\documentclass[preprint2]{aastex}
\tightenlines 
\doublespace
\onecolumn

\newcommand{\etal}{{et~al.~}}

\newcommand{\tmb}{\mbox{$\rm T_{\rm MB}$}}

\newcommand{\tp}{\mbox{$\rm T_p$}}
\newcommand{\lmax}{\mbox{$\rm l_{max}$}}
\newcommand{\lmin}{\mbox{$\rm l_{min}$}}
\newcommand{\co}{\mbox{$^{12}$CO}}
\newcommand{\coa}{\mbox{$^{13}$CO}}
\newcommand{\lco}{\mbox{L$_{CO}$}}
\newcommand{\mco}{\mbox{M$_{CO}$}}
\newcommand{\cc}{\mbox{${\rm cm}^{-3}$}}

\newcommand{\kms}{\mbox{${\rm km~s}^{-1}$}}
\newcommand{\fwhm}{\mbox{$\delta v$}}
\newcommand{\vlsr}{\mbox{$\rm V_{\rm LSR}$}}

\newcommand{\htwo}{\mbox{${\rm H}_2$}}

\newcommand{\msun}{\mbox{$M_\odot$}}

\begin{document}
\title{The Equilibrium State of Molecular Regions in the Outer Galaxy}
\author{Mark H. Heyer}
\affil{Five College Radio Astronomy Observatory and 
Department of Astronomy,\\
Lederle Graduate Research Tower,
University of Massachusetts,
Amherst, MA 01003}
\author{John M. Carpenter}
\affil{Department of Astronomy, 
California Institute of Technology, \\
Mail Stop 105-24, Pasadena, CA 91125}
\author{Ronald L. Snell}
\affil{Five College Radio Astronomy Observatory and 
Department of Astronomy,\\
Lederle Graduate Research Tower,
University of Massachusetts,
Amherst, MA 01003}
\vskip .75in

\parindent=20pt

\begin{abstract}
A summary of global properties and an evaluation of the 
equilibrium state of molecular regions in the 
outer Galaxy are presented from the decomposition 
of the FCRAO Outer Galaxy Survey and targeted
\co\ and \coa\ observations of four giant molecular 
cloud complexes.  The ensemble of identified objects 
includes both small, isolated clouds and clumps within 
larger cloud complexes. 
The  \co\ luminosity function and size distribution of a subsample 
of objects with well defined distances are 
determined such that 
$ { {{\Delta}N} \over {{\Delta}L_{CO}} } = (3{\times}10^4) L_{CO}^{-1.80\pm0.03}$
and 
$ { {{\Delta}N} \over {{\Delta}r_e} } = (1.7{\times}10^4) r_e^{-3.2\pm0.1} $.
\co\ velocity dispersions show little variation
with cloud sizes for radii less than 10 pc.
It is demonstrated that the internal motions 
of regions with \mco=$X_{CO} L_{CO} >$ 10$^4$ \msun\
are bound by self gravity, yet,  the constituent clumps of
cloud complexes and isolated molecular clouds with \mco $<$ 10$^3$ \msun\
are
not in self gravitational equilibrium.   The required external 
pressures to maintain the equilibrium of this population
are (1-2)$\times$10$^4$ cm$^{-3}$-K.   
\end{abstract}

\parindent=20pt
\parindent=20pt

\section{Introduction} \label{INTRO}

Molecular regions in the Galaxy exist within a wide range of environmental
conditions.  There are massive giant molecular 
clouds near the Galactic Center with large mean densities
(Bally \etal 1988), highly excited 
molecular gas associated with ionization fronts and 
supernova remnants (Elmegreen \& Lada
1977), quiescent clouds and globules (Clemens \& Barvainis 1988), 
and diffuse, high latitude clouds
with low column densities in 
the solar neighborhood (Magnani, Blitz, \& Mundy 1985).
In addition to local sources of perturbation, the molecular gas
responds to large scale variations in the Galaxy such as spiral 
potentials and the surface density of stars
and gas
(Elmegreen 1989).
These different environments and conditions regulate the stability
of the gas and therefore, modulate the formation of stars. 
Therefore, it is important to evaluate the molecular gas properties
over a wide range of environments. 

A general description of the molecular interstellar medium requires 
surveys of molecular line emission over large volumes of the 
Galaxy with high angular and spectral resolution and sampling.  Such surveys 
provide a census of the molecular gas without an undue bias toward bright
emission or association with active sites of star formation.  The 
subsequent large number of molecular regions identified in wide field
surveys enable a statistical evaluation of gas properties and 
classification with respect to the local environment. 
There have been several important wide field surveys 
of CO emission from the Galaxy.  The large scale distribution
of molecular gas in the Milky Way has been determined from the
combined North-South surveys summarized by Dame \etal (1987).
However, the large effective beam size limits the description 
of gas properties to 
the largest giant molecular cloud complexes.   
The Massachusetts-Stony Brook Survey imaged the inner
Galaxy with an effective resolution of 3\arcmin\ (Sanders \etal 
1985).  Analysis of the data by 
Solomon \etal (1987) and Scoville \etal (1987) identified a
number of giant molecular clouds and cloud  complexes.
Many of the accepted characteristics of the 
molecular interstellar medium are derived from these studies. 
These include the self gravitational equilibrium state of the 
giant molecular clouds and the relationship between the velocity 
dispersion and size of the cloud. 
However, cloud properties determined from inner Galaxy surveys
are compromised due to the high degree of confusion along the line of 
sight due to velocity crowding 
which precludes a complete accounting of the 
emission (Lizst \& Burton 1981). To reduce the blending of emission
from unrelated clouds, molecular regions 
are identified as high temperature 
isophotes within the longitude-latitude-\vlsr\ volume.
While the 
large temperature thresholds reduce the 
confusion along the line of sight, these necessarily bias the 
resultant cloud catalogs to the warmest, densest regions within the 
molecular interstellar medium.  \co\ and \coa\ imaging observations of 
targeted clouds demonstrate that most of the molecular mass resides within the 
extended, low column density lines of sight (Carpenter, Snell \& Schloerb 1995;
Heyer, Carpenter \& Ladd 1996).   Such low column 
density regions in the inner Galaxy
are
simply  not accessible for the  analysis of cloud properties.

In contrast, the outer Galaxy provides a less confusing view 
of the molecular interstellar medium. 
Beyond the solar circle, 
there is no blending of emission from widely separated clouds along the 
line of sight.  Therefore, molecular 
regions can be identified at lower gas column densities 
from which more representative global properties can be derived.
This property has been exploited in a series of investigations 
by Brand \& Wouterloot
(1994, 1995).
While these studies provide a  sensitive, high resolution
perspective of individual clouds and the distribution of molecular regions in 
the far outer Galaxy,  the results are necessarily biased toward 
clouds associated with star formation.

The FCRAO CO Survey of the outer Galaxy provides an
opportunity to study the equilibrium state of molecular clouds
under varying conditions (Heyer \etal 1998).  The survey 
searched for $^{12}$CO J=1-0 emission within a 
330 deg$^2$ field sampled every 50\arcsec\ with a FWHM beam size of 
45\arcsec.  The \vlsr\ range is -153 to 40 \kms\ sampled 
every 0.81 \kms\ with a resolution of 0.98 \kms. The median main beam 
sensitivity (1$\sigma$) per channel
is 0.9 K.
In this contribution, we present results from a decomposition of the 
outer Galaxy Survey into discrete objects.  \co\ luminosity, size, and 
line width 
distributions are determined from the ensemble of identified 
objects located in the Perseus arm and far outer Galaxy.
In $\S$3, we reexamine the Larson scaling relationships 
with the cloud catalog extracted from the Survey and 
with clumps from a similar decomposition of \co\ and \coa\
observations of several targeted giant molecular clouds.  
\section{Results}

To isolate discrete regions of CO emission from the large data cube,
we have adopted the definition of a molecular cloud used 
by previous investigations (Solomon \etal 1987; Scoville \etal 1987;
Sodroski 1991).  That is, a discrete molecular region is identified as a closed
topological surface within the $l-b-V_{LSR}$ data cube 
at a given threshold of antenna 
temperature.  
In this study, the limiting threshold is 1.4 K (main beam temperature scale)
or 1.5$\sigma$ where $\sigma$ 
is the median rms value of antenna temperatures in the Survey (Heyer \etal 1998).
The threshold value is sufficiently low to provide a more complete 
accounting of the flux within the data cube as compared to the inner 
Galaxy surveys, while large enough to exclude misidentifications of
molecular regions due to statistical noise.   Detailed descriptions of 
the cloud decomposition and the calculation of cloud properties
are provided in Appendix A.

\subsection{The Outer Galaxy Survey Cloud Catalog}

The decomposition of the FCRAO CO Survey of the Outer Galaxy at 
a limiting threshold of \tmb=1.4 K yields 10156 objects.  
Each object is
described by position centroids ($l_c,b_c,v_c$), velocity width, 
\fwhm, a kinematic distance, D, assuming purely circular motions and 
a flat rotation curve
\footnote{$\Theta_\circ$=220 \kms; $R_\circ$=8.5 kpc},
Galactocentric radius, R$_{gal}$, z height, CO luminosity, \lco, and 
a peak antenna temperature within the surface, \tp.  The geometry of an object 
is described by 
major and minor axis diameters, \lmax, \lmin,
and a position angle, $\theta$,
 of the major axis with respect to the 
Galactic plane (see Appendix A).  All sizes are derived assuming a 
kinematic distance to the object.  The derived properties of the identified 
objects are listed in Table~1.  Objects in the catalog are named
HCS followed by the sequential catalog number.

For much of the subsequent analysis, we exclude molecular regions 
with $v_c >$-20 \kms\ since the kinematic distances are not 
sufficiently accurate for such local emission. 
The gas in the outer Galaxy
is known to exhibit large deviations from 
circular motions (Brand \& Blitz 1993).
For example, the IC1805 OB cluster has a spectroscopic distance of 2.35 kpc
which corresponds to a circular velocity of -20 \kms.
The bulk of the CO emission occurs at \vlsr\ of -40 to -50 \kms\
with a kinematic distance of 4-5 kpc.  Such discrepancies have been
attributed to streaming motions of the gas in response to the 
spiral  potential or a triaxial spheroid (Blitz \& Spergel 1991).
In the Survey field, kinematic distances can be larger than 
the spectroscopic distances by factors of 2.
Therefore, in some cases, CO luminosities and inferred molecular hydrogen
masses may {\it overestimate} the true values by factors of 3 to 4.
The restricted subset with $v_c <$ -20 \kms\ 
is comprised of 3901 objects with kinematic distances greater than 
$\sim$2 kpc and galactocentric radii greater than $\sim$9.5 kpc.  

\begin{figure}
\plotone{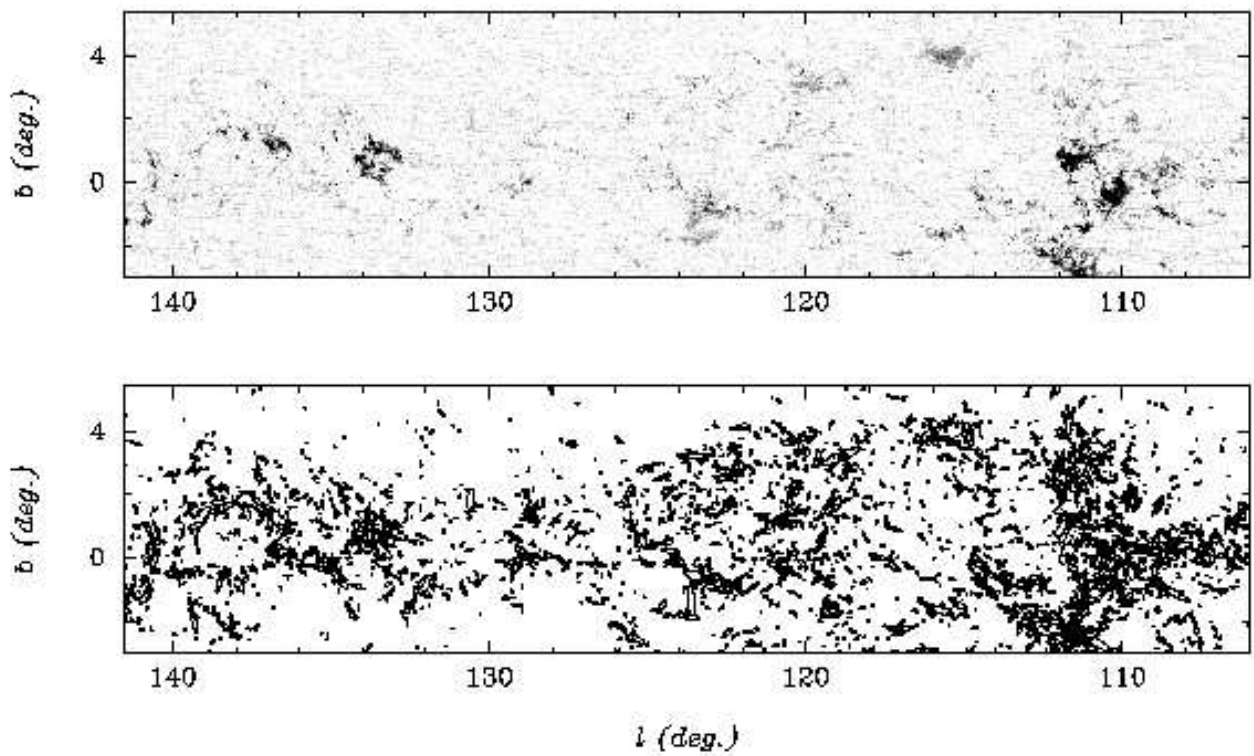}
\caption{(top) An image of $^{12}$CO J=1-0 integrated intensity
over the velocity interval -110 to -20 \kms from the FCRAO CO Survey
of the outer Galaxy (Heyer \etal 1998).   The halftone ranges
from 0 (white) to 20 (black) K km s$^{-1}$.
(bottom) The positions, sizes, and orientations of identified 
objects approximated as ellipses. }
\label{survey.clouds}
\end {figure}

The decomposition extracts both isolated molecular clouds and
clumps within larger cloud complexes which are identified separately 
due to the spatial and 
kinematic inhomogeneity of the molecular interstellar medium.  
Any element within the data cube can only be assigned to a
singular object.
Figure~\ref{survey.clouds} shows an image of integrated intensity
over the velocity range -110 to -20 \kms\
and the ensemble of identified objects whose positions, inferred sizes
and orientations are represented as ellipses with parameters \lmax, \lmin,
and 
$\theta$.

\subsection{Selection Effects}

Prior to the examination of cloud properties, it is necessary to 
establish known selection effects.
The primary selection effects are due to 
the limited spectral resolution of the observations,  
the antenna temperature threshold, and the requirement that an object 
be comprised of at least 5 spatial pixels (see Appendix A).
The antenna temperatures of 
two contiguous spectroscopic channels 
are required to be larger than the main beam temperature 
threshold of 1.4 K in order for these
channels to be 
associated with an object.   
Therefore, 
the cloud catalog does not include 
molecular regions with narrow ($\sigma_v$ $<$ 0.4 \kms) velocity dispersions.  
In Appendix B, we evaluate 
the sensitivity and accuracy of the method  from the cloud decomposition of 
simple cloud models with varying signal to noise and 
line width.  
The decomposition can
recover velocity dispersions as small as 
0.65 \kms\ to within 10\% accuracy for signal to noise ratios greater than 4.
The results are
important not only to gauge the selection of objects but to evaluate
the cloud scaling laws 
discussed in $\S$3.

Given the antenna temperature threshold of 1.4 K and assuming that \co\
is universally optically thick, the decomposition can not 
recover regions with extreme subthermal excitation conditions 
averaged over
the angular extent of the cloud.
The intensity threshold and 
spectroscopic channel requirements correspond to an H$_2$ column density of 
4$\times$10$^{20}$ cm$^{-2}$ (see $\S$2.4) at which self shielding of CO molecules may not 
be effective (van Dishoeck \& Black 1988).  Therefore, the edges defined by this
threshold likely correspond to a photodissociation boundary of CO.
The distribution of \htwo\ gas is likely more extended than the CO.
Independent of excitation, the catalog would not include 
 small, compact clouds
with mean angular radii less than $\sim$2 arcminutes due to the 
requirement that the cloud is comprised of at least five pixels. 

\subsection{Detection and Completeness Limits for the Sample}

\begin{figure}
\plotone{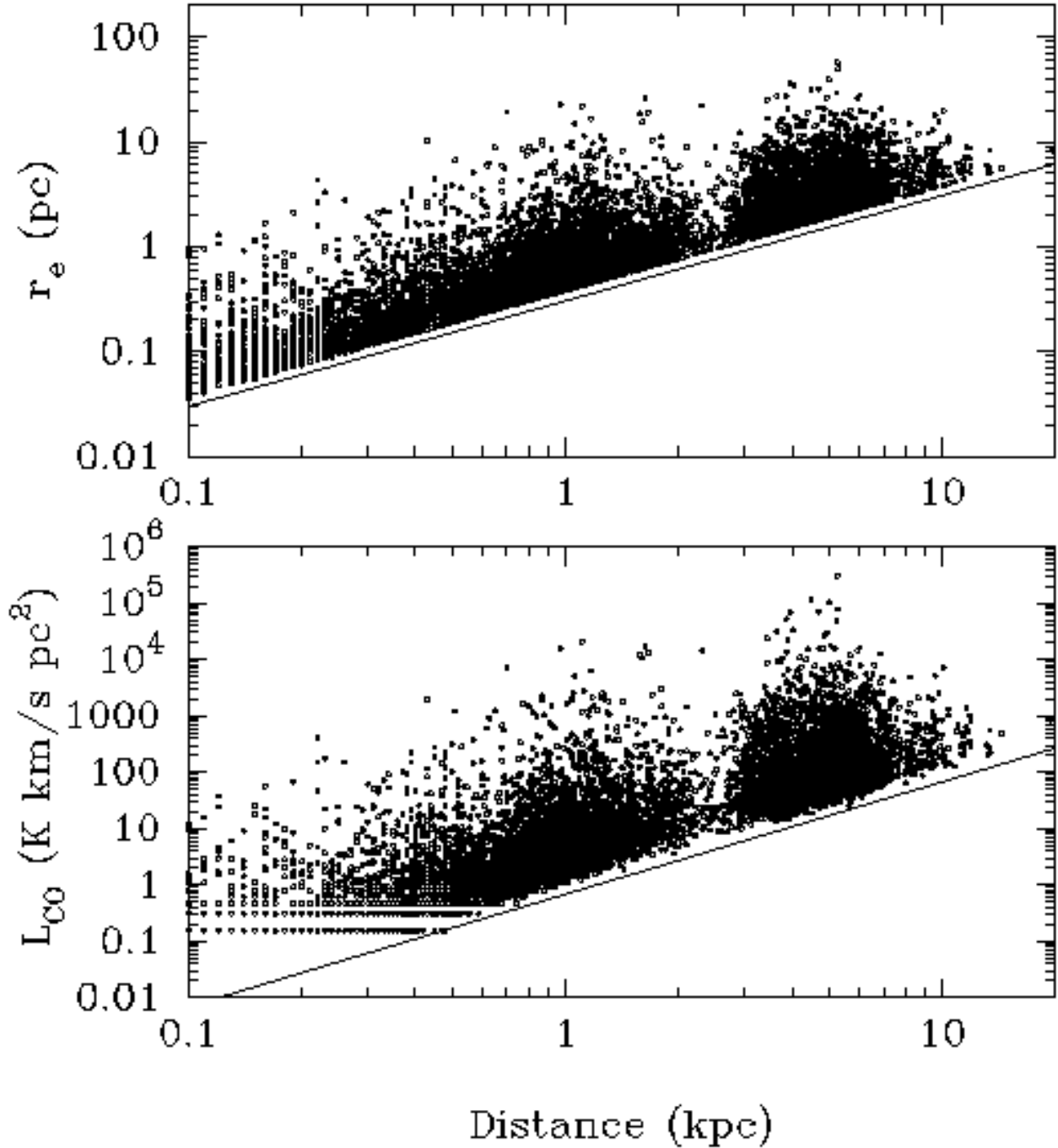}
\caption{(top)The effective radius of identified objects as a function of 
kinematic distance.  The lower envelope of points is due 
to the requirement that an object must be comprised of at least 
five spatial pixels.
(bottom) \lco\ as a function of distance.  The solid line shows the 
minimum CO luminosity of objects in which the antenna temperature of 
all associated 
positions and spectroscopic channels is 1.4 K. }
\label{complete}
\end {figure}

Due to the prescribed definition of a cloud, there are limits to
physical quantities such as size and CO luminosity below 
which the cloud catalog is insensitive and incomplete.
Figure~\ref{complete}
shows the effective cloud radius, $r_{e} = 0.5 \sqrt{\lmax\lmin}$ and 
CO luminosity as functions
of the kinematic distance for all identified objects.   The lower 
envelope of points corresponds to the minimum effective
size of a cloud,
$$ r_{e}^{min}(D) = 10^3 D \sqrt{\Omega_s N_p/\pi}  \;\;\;pc \eqno(2.1)$$
where $\Omega_s$ is the solid angle per pixel, D is the distance to the 
object in kpc, and $N_p$ is the minimum number of pixels per object.  For
this decomposition, $N_p=5$ so that $r_{e}^{min}=0.31D$ pc.  
Similarly, the minimum CO luminosity is, 
$$L_{CO}^{min}(D) = N_p N_c T_{th} dv \Omega_s D^2 \;\;\;K\;km\;s^{-1}\;pc^2
 \eqno(2.2)$$
where $N_c=2$, is the minimum number of velocity channels, $dv$=0.81 \kms,
 is
the 
spectroscopic channel width, and $T_{th}$=1.4 K is the main beam 
antenna temperature
threshold. $L_{CO}^{min}(D)$ is 
shown as the solid line in Figure~\ref{complete}.  At a distance of 10 kpc,
the detection limit is 67 $K\;km\;s^{-1}\;pc^2$.
While $L_{CO}^{min}(D)$ is the detection limit at a given distance, D,
the completeness limit is higher than this value since the noise of the 
data contributes to the measured luminosity.  
The completeness limit, $L_{CO}^c$ is defined in this study at 
the 5$\sigma$ confidence as
$$L_{CO}^{c}(D) = L_{CO}^{min}+5\sigma(L_{CO}) \eqno(2.3) $$
where 
$$\sigma(L_{CO}) = \sigma dv \Omega_s D^2 \sqrt{N_cN_p} \;\;\;K\;km\;s^{-1}\;pc^2
 \eqno(2.4)$$
and $\sigma$=0.93 K is the median rms temperature for channels 
with no emission (Heyer \etal 1998).  At 10 kpc, the cloud catalog
is complete for CO luminosities $>$ 138 $K\;km\;s^{-1}\;pc^2$.
This completeness limit needs
to be considered when calculating power law descriptions to the 
CO luminosity function in $\S$2.5.1. 

\subsection{ Application of the CO to H$_2$ Conversion Factor}

Whenever possible, descriptions of observable 
quantities are presented with few or no assumptions.  However, for analyses
described in $\S$3.2 and $\S$4.1, it is necessary to derive
total gas column densities and masses.  In this investigation
for which only \co\ observations are available, 
\htwo\ column densities  are derived using the CO to \htwo\ conversion factor
determined from $\gamma$-ray measurements such that 
$$ N(H_2) = 1.9{\times}10^{20} W_{CO} \;\;\; cm^{-2} \eqno(2.5)  $$
where $W_{CO}$ is the \co\ integrated intensity in K km s$^{-1}$ 
for a given line 
of sight (Strong \& Mattox 1996).  Summing over the projected area 
of the cloud, this corresponds to a conversion from CO luminosity, \lco,
in $K\;km\;s^{-1}\;pc^2$ to the total molecular mass, \mco\
 of the identified object,
$$ M_{CO} = 4.1 \biggl({{L_{CO}}\over{K\;km\;s^{-1}\;pc^2}}\biggr)  \;\; M_\odot  \eqno (2.6) $$
which includes the factor 1.36 to account for the abundances of heavier
elements
(Hildebrand 1983).

The dimensional justification for a constant 
conversion factor is summarized by Dickman, Snell,\& Schloerb (1986).
The CO luminosity is the integral of the antenna temperature over all
velocities and area of the cloud,
$$ L_{CO}= {\int}dA(l,b){\int}dvT(l,b,v) \;\;\; K\;km\;s^{-1}\;pc^2 $$
$$ L_{CO} \approx \pi r_e^2 <T>{\delta}v \;\;\;\; K\;km\;s^{-1}\;pc^2 \eqno(2.7) $$
where dA is the projected area of a pixel in pc$^2$, 
$<T>$ is the mean brightness temperature over the projected area and
velocity interval, ${\delta}v$ is the full width half maximum line width
in $km s^{-1}$, and 
$r_{e}$ is the effective radius in pc.  The gravitational parameter, $\alpha_G$,
for a spherically symmetric, uniform density cloud is 
$$ \alpha_G \approx { {5\sigma_v^2r_e} \over {GM} } \eqno(2.8) $$
where $\sigma_v$ is is the velocity dispersion of the object.
Solving for ${\delta}v=\sigma_v(8ln2)^{1/2}$ 
in equation 2.8 and substituting 
into equation 2.7,
$$ L_{CO} = 1.05 \pi r_e^2 <T> \biggl({ {GM\alpha_G} \over {r_e} }\biggr)^{1/2} \eqno(2.9) $$
Dividing equation 2.9 into $M=4/3\pi r_e^3 <\rho>$, where $<\rho>$ is the mean density
of the cloud yields $X_{CO}$,
$$ X_{CO} = {{M}\over{L_{CO}}} $$
$$ X_{CO} = 0.62 <T>^{-1} (G{\alpha_G})^{-1/2} <\rho>^{1/2} \eqno(2.10)$$
$$ X_{CO} = 4.1 \biggl({{<T>}\over{5 K}}\biggr)^{-1} \biggl({ {<n>}\over{100 cm^{-3}}}\biggr)^{1/2}{\alpha_G}^{-1/2}  \eqno(2.11) $$
where $<n>={\mu}m_{H_2}<\rho>$.
The conventional assumption is that clouds are self gravitational 
($\alpha_G \approx 1$)
 and the 
mean temperature and density do not vary from cloud to cloud such that 
$X_{CO}$ is constant.   However,
if clouds are not gravitationally bound ($\alpha_G >> 1$), 
then the appropriate value of $X_{CO}$ decreases with 
respect to the value for a self gravitating cloud.
Therefore, by applying a constant, universal value of $X_{CO}$,
under the assumption that 
$\alpha_G=1$, the resultant \htwo\ mass {\it overestimates} the true mass 
of a cloud with $\alpha_G >> 1$.

\subsection{Distributions of Measured Properties} 
The large number of objects identified in the decomposition of the FCRAO
CO Survey of the Outer Galaxy enables a detailed examination of the 
CO luminosity function, the size spectrum, and line
width distribution of molecular regions. 

\subsubsection{CO Luminosity Function} 
The mass spectrum of clouds, N(m)dm, within the Galaxy and clumps
within molecular cloud complexes is used 
as a diagnostic to cloud formation and fragmentation processes (Kwan 1979);
a signature of a hierarchical interstellar medium (Elmegreen \& Falgarone 1996;
Stutzki \etal 1998);
and a guide to the initial stellar mass function 
(Zinnecker \etal 1993).
It is typically quoted as a
differential distribution which is described by a power law,
$$ { {dN} \over {dM} } \propto M^{-{\alpha_M}} \;\;\;\;\; \eqno(2.12)$$
Values for $\alpha_M$ are similar for distributions
describing clouds and cloud complexes in  the Galaxy or clumps within
clouds.
Kramer \etal (1998) summarize the mass spectra of clumps 
within several cloud complexes.  They find that $\alpha_M$ ranges
from 1.6 to 1.8 over a large range of mass scales.  
From a sample of giant molecular clouds identified within coarsely
sampled surveys of CO emission from the inner Galaxy,
Sanders, Scoville \& Solomon (1985) and Solomon \etal (1987) derive 
$\alpha_M=1.5$.   
For these 
Galactic surveys, the mass of a cloud
is determined from the virial mass and the assumption of self gravitational
equilibrium.

\begin {figure}
\plotone{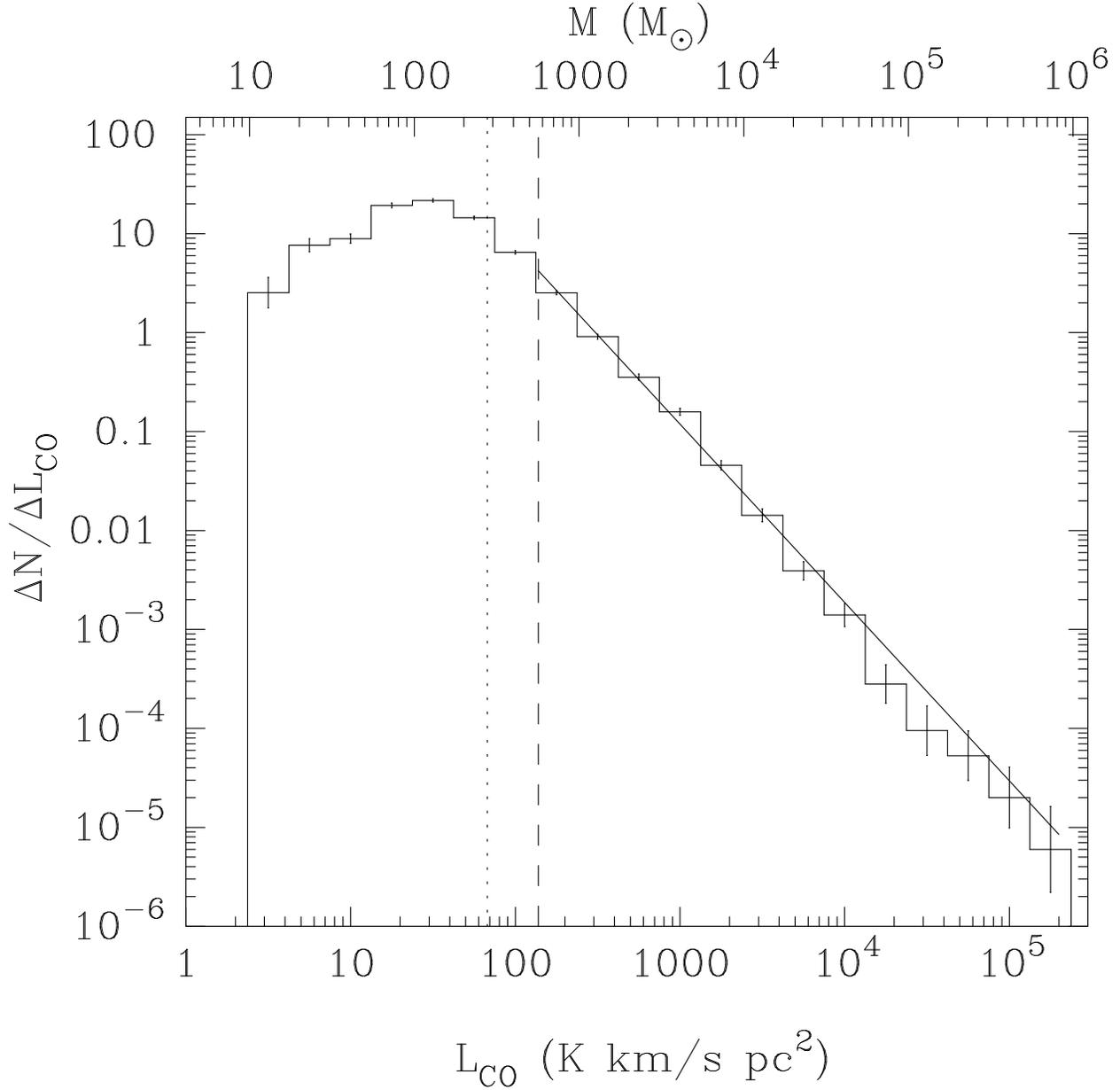}
\caption{The CO luminosity function, ${\Delta}N/{\Delta}L_{CO}$ for 3901 identified objects.
The top x coordinate shows the corresponding mass scales assuming a constant 
CO to H$_2$ conversion factor.  
The vertical dotted 
line denotes the detection limit of \lco\ and the vertical dashed 
line marks the completeness
limit of \lco\ at a distance of 10 kpc.  The power law fit to bins
above 
the completion limit (solid line) is ${\Delta}N/{\Delta}L_{CO}$ $\propto$ \lco$^{-1.80\pm0.03}$. }
\label{lco-function}
\end {figure}

 Figure~\ref{lco-function} shows the differential luminosity function,
${\Delta}N/{\Delta}L_{CO}$,  in equally spaced, logarithmic bins for objects with 
\vlsr$<$-20 \kms.  The corresponding mass for a given \lco\ is shown
along the top x coordinate assuming a constant CO to H$_2$ conversion factor
(see $\S$2.4).  
The detection limit of the sample at a distance of 10 kpc is 
67 $K\;km\;s^{-1}\;pc^2$.
The sample is complete to a limiting value of 
138 $K\;km\;s^{-1}\;pc^2$ at a distance of 10 kpc.
A power law is fit to the bins with \lco\ greater 
than this completeness limit and weighted by $\sqrt{N}/{\Delta}L_{CO}$
where $N$ is the 
number of objects in each bin such that 
$$ { {{\Delta}N} \over {{\Delta}L_{CO}} } = (3.1{\times}10^4) L_{CO}^{-1.80\pm0.03} 
(K\;km\;s^{-1}\;pc^2)^{-1} \eqno(2.13) $$
The value of the exponent is steeper than the value derived by
Brand \& Wouterloot (1995) for a sample of outer Galaxy clouds
for which $\alpha_M$=1.62.  
This may be partly due to the improved statistics 
given the large number of 
objects and a lower luminosity 
completeness limit.  Nevertheless, the value of the exponent
is less than the critical value of 2, at which there are equal integrated 
luminosities (masses) over any logarithmic range of luminosity.
Therefore, most of the observed flux is contributed by the 
most luminous objects.  For example,
50\% of the luminosity integrated over all identified objects (3901)
comes from 35 clouds with \lco $>$ 7800 $K\;km\;s^{-1}\;pc^2$ 
and 90\% is contributed by 930 clouds with \lco $>$ 270 $K\;km\;s^{-1}\;pc^2$. 

\subsubsection{Cloud Size Distribution} 
\begin {figure}
\plotone{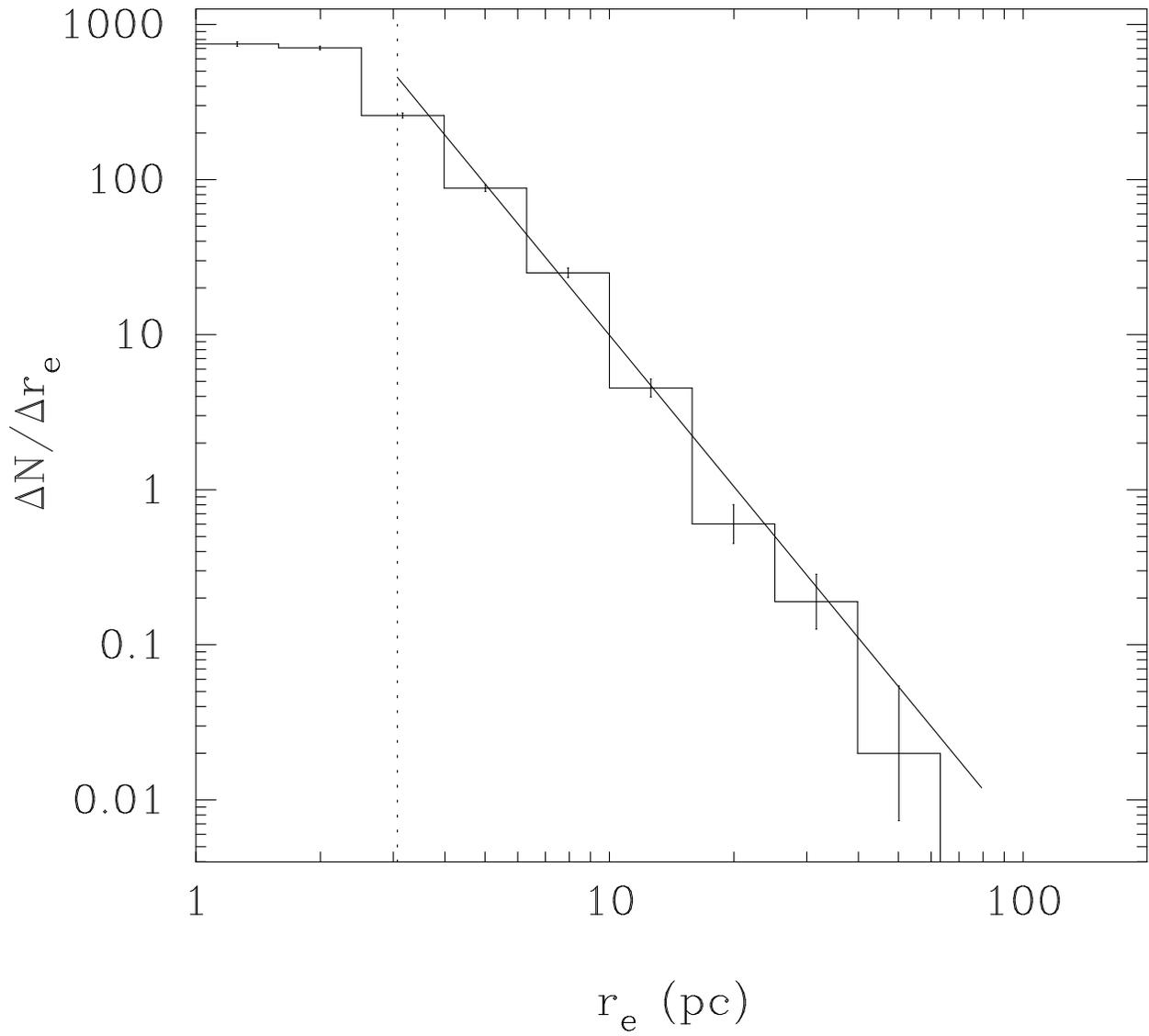}
\caption{The size distribution, ${\Delta}N/{\Delta}r_e$, for the 
identified objects.
The 
detection 
limit of r$_e$ at a distance of 10 kpc is 3.1 pc.  The power law fit to bins
above 
this limit (solid line) is ${\Delta}N/{\Delta}r_e$ $\propto$ r$_e^{-3.2\pm0.1}$. }
\label{size-distr}
\end {figure}

The size distribution of identified objects provides an additional 
measure of mean cloud properties.  The differential 
size distribution, ${\Delta}N/{\Delta}r_{e}$, 
is presented in Figure~\ref{size-distr}.
The detection limit at a distance of 10 kpc is shown as the 
vertical dotted line.  A power law fit to bins above 
this limit and weighted by $\sqrt{N}/{\Delta}r_e$, 
yields the relation, 
$$ { {{\Delta}N} \over {{\Delta}r_e} } = (1.7{\times}10^4) r_e^{-3.2\pm0.1} \;\;\; pc^{-1}
\eqno(2.14) $$
The size spectrum for the outer Galaxy clouds is similar to that 
derived from inner Galaxy surveys (Solomon \etal 1987) and targeted
molecular regions over a more limited range of cloud size
(Elmegreen \& Falgarone 1996).  

\subsubsection{Line Width Distributions}
\begin {figure}
\plotone{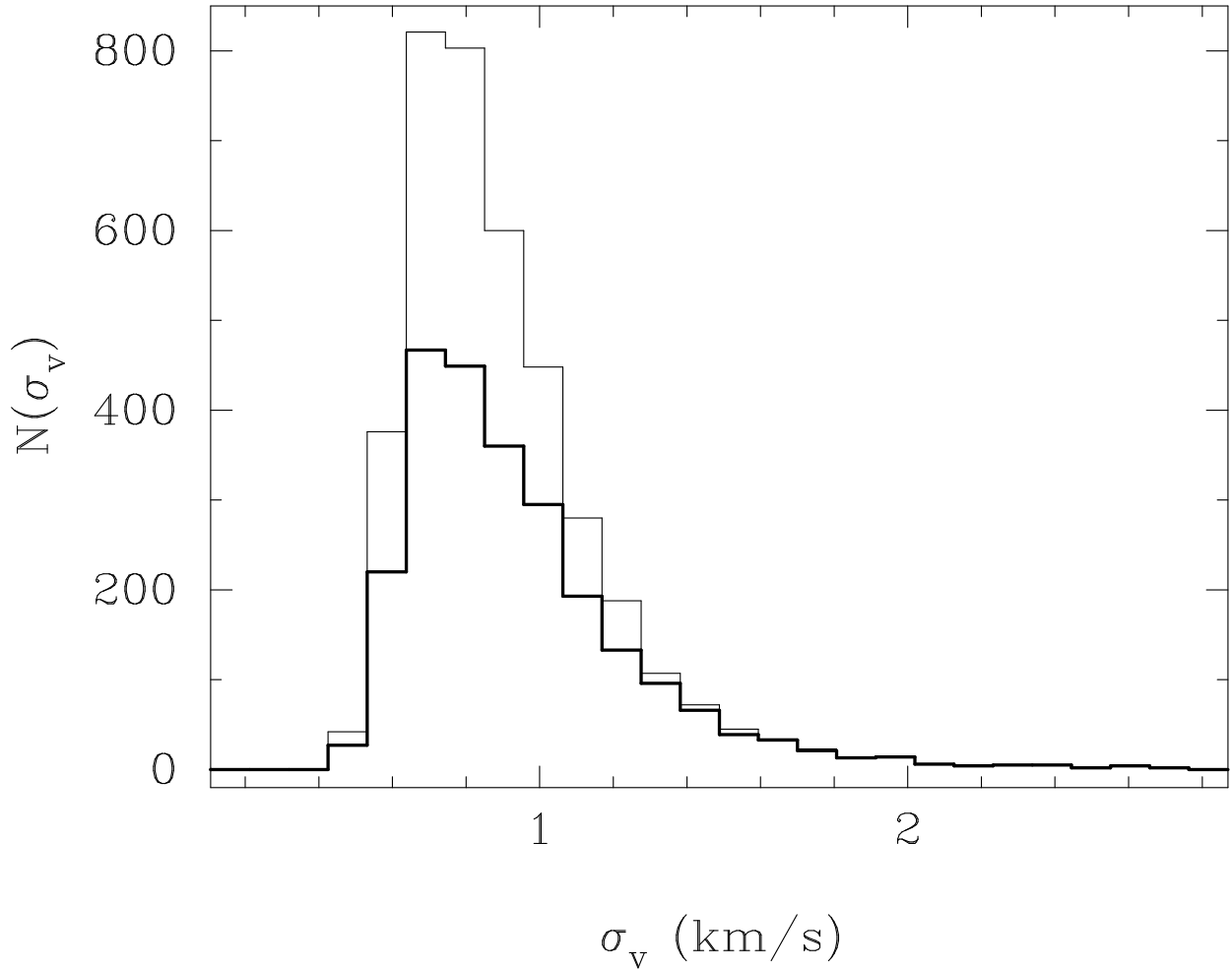}
\caption{The number distribution, N($\sigma_v$) of measured velocity 
dispersion. 
The heavy solid shows the distribution of 
velocity dispersion for objects with a peak antenna temperature 
$>$ 3.5 K for which the velocity dispersions are well determined.
}
\label{vw-distr}
\end {figure}

Each object is characterized by a velocity dispersion,
$\sigma_v={\delta}v/(8ln2)^{1/2}$, 
derived from the summed spectrum of all constituent pixels.
This measure includes both line of sight motions as
may be inferred from the mean line width of each profile 
and projected variations of the centroid velocities.  Therefore, 
it accounts for all of the measured kinetic energy within a cloud
generated by turbulence, rotation, expansion, and other dynamical processes.
The distribution of velocity dispersions, $N(\sigma_v)$, 
is shown in Figure~\ref{vw-distr}.  
The 
decomposition is not sensitive to small line width regions and provides
accurate measures for $\sigma_v$ $>$ 0.64 \kms.
The distribution of velocity dispersions for clouds with large
peak temperatures ($>$ 3.5 K) and presumably more accurate values of
$\sigma_v$, 
shows the same shape and mean value as the distribution for all
objects.  The measured distribution of line
width for all objects is not significantly biased by signal 
to noise or errors in measuring $\sigma_v$.

\section{Cloud Scaling Relationships}
With the statistics of individual properties established in the
preceding sections, we now examine the relationships between various 
cloud properties.   These relationships are motivated by the scaling
laws initially identified by Larson (1981) and reexamined
by many subsequent studies with varying results.  
These scaling laws describe 1) a power 
law relationship between the velocity dispersion 
and size of a cloud;
2) a linear correlation between the measured and virial mass 
of clouds;
and 3) an inverse relationship between 
mean density and size.
The three relationships are algebraically 
coupled such that the validity 
of any two of these laws necessarily  implies the third (Larson 1981).

\subsection{Velocity Dispersion - Size Relationship}
The velocity dispersion, $\sigma_v$, provides a 
measure of the total kinetic energy in the cloud inclusive of 
thermal, turbulent, rotational, and expanding motions.  
 A scaling relationship between the velocity dispersion and the 
size of a cloud was initially identified by Larson (1981) using 
data taken from the literature.  
A recent compilation of data
from many studies using several different molecular line tracers
demonstrates this correlation of velocity dispersion with size
within 4 orders of magnitude in size scale (Falgarone 1996).
The origin of 
this relationship has been attributed to turbulence (Larson 1981;
Myers 1983), or simply, a consequence of gravitational 
equilibrium and 
constant gas column density which are limited by observational 
selection effects (Scalo 1990).  

It is important to distinguish the relationship derived using multitracer
observations from that determined from a single gas tracer (Goodman \etal 1998).
The excitation requirements for a given molecule determine the angular 
extent over which any object can be identified.  
Multitracer observations sample different density regimes which correspond to 
distinct, but nested, volumes of material.
In this way, a larger dynamic range of 
sizes is probed than can be sampled by any single gas tracer.
The correlation between velocity dispersion and size has been established for
single gas tracers but over a more limited range of sizes and larger
intrinsic scatter (Larson 1981; Dame \etal 1987; Solomon \etal 1987).  
These single tracer relationships examine the variation of the velocity
dispersion with size within a more limited range of density.

\begin {figure}
\plotone{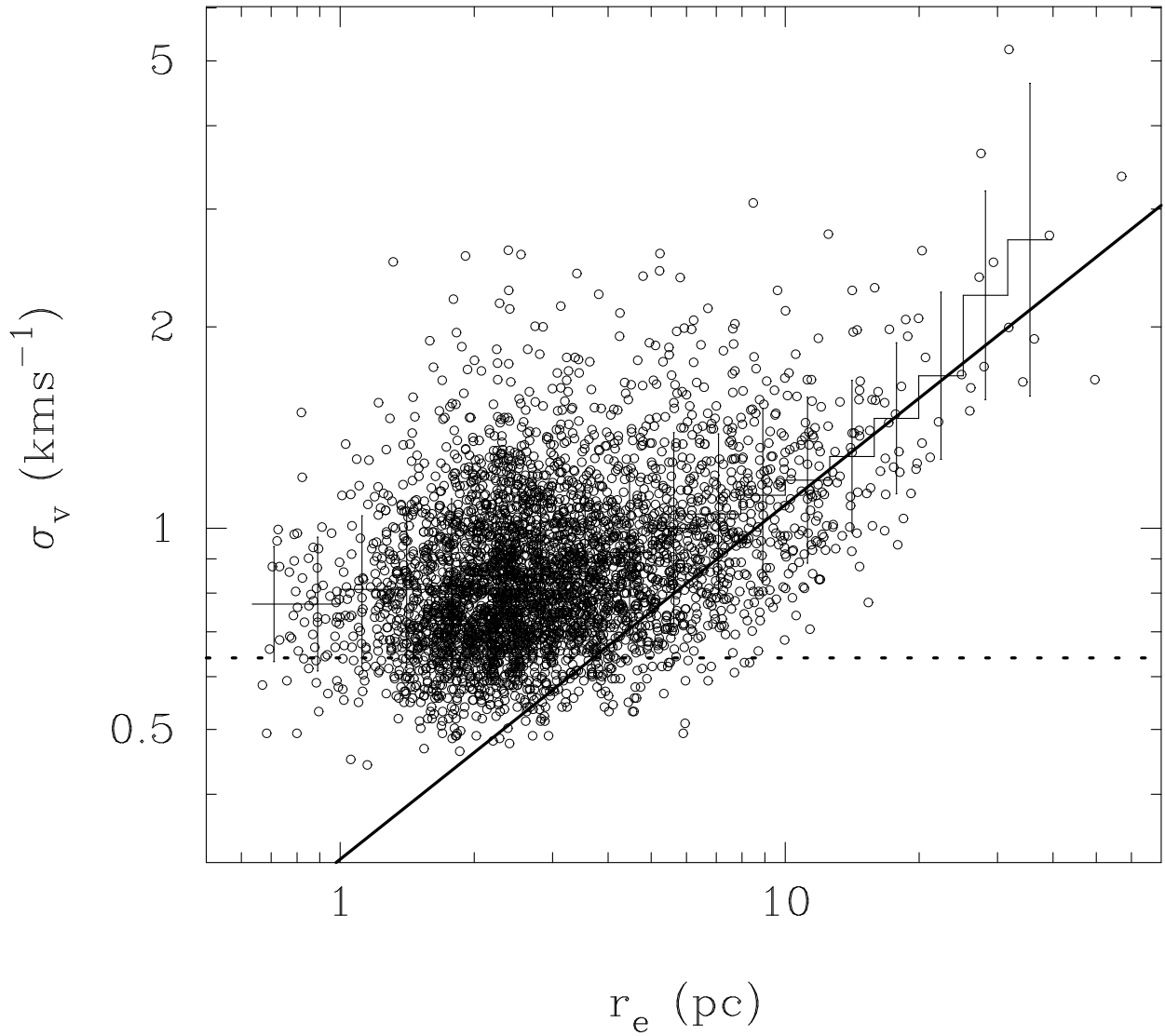}
\caption{The variation of measured velocity dispersion, $\sigma_v$,
and effective size, $r_e$.  The light line shows the mean value
within logarithmic bins of $r_e$ and the error bars reflect the 
dispersion of values about the mean in each bin.  The heavy line
shows the power law fit to the clouds with $r_e >$ 9 pc.  The 
slope of the power law is similar to that found by Solomon \etal (1987).
The horizontal dashed line shows the velocity dispersion to
which the measured values are accurate to within 15\%.
}
\label{vw-r}
\end {figure}

Figure~\ref{vw-r} presents the variation of velocity dispersion 
with the effective radius, $r_e$, for the ensemble of clouds in this study. To 
more effectively consolidate the information, the 
mean velocity dispersion is calculated within binned cloud radii. 
For objects with sizes greater than $\sim$7 pc, there 
is a tendency for increasing 
velocity dispersion with size.  The slope of the power law fit to 
objects with radii greater than 
9 pc is  $\sim$0.5 and similar to that derived by Solomon \etal (1987).
However, the binned 
values show little systemic variation of the velocity dispersion with size for 
$r_e <$ 7 pc.  
The apparent flattening of the relationship for small clouds is not 
an artifact of our cloud definition 
since it occurs at a 
velocity dispersion for which our method is reasonably
accurate (see Appendix B).
A limited number of 
followup observations with much higher spectral resolution 
of narrow line width clouds identified in the 
catalog show comparable velocity dispersions (see Appendix B).
A population of 
small clouds with line widths below our threshold for cloud identification
which do follow the standard relationship can not be excluded. 
However, our observations have identified many small clouds 
with line widths in excess of the extrapolated size line-width
relationship of Solomon \etal (1987).
This result does not dismiss the velocity dispersion-size 
relationship determined from multitracer observations. 
Narrow line width regions within molecular clouds 
are identified from tracers of high density
gas (NH$_3$, CS, HCN) and these often follow the 
conventional scaling law (Myers 1983). Such regions are not readily 
identified by \co\ or \coa\ emission.
Previous 
CO studies which have identified a size line width scaling law have been 
limited to large, self gravitating cloud complexes with 
masses greater than 10$^4$ \msun\ (Solomon \etal 1987, Scoville \etal 
1987).  
The near constant velocity dispersion with size 
for the small cloud or clump
population may reflect a different dynamical state than the 
larger giant molecular cloud complexes (see $\S$3.2).

\subsection{Equilibrium of Molecular Regions}
\subsubsection{Survey Clouds}

To evaluate the role of self gravity in the equilibrium 
of the identified molecular regions, we determine the 
magnitude of the virial mass with respect to the measured 
mass of the object derived from the CO luminosity. 
Following Bertoldi \& McKee (1992), the 
 virial mass, $M_{vir}$, is calculated  from the measured cloud parameters, 
$$ M_{vir} = {{a_3}\over{a_1 a_2}} {{5\sigma_v^2 r_e}\over{G}} \eqno(3.1) $$
where $\sigma_v={\delta}v/(8ln2)^{1/2}$ is the one dimensional velocity 
dispersion.
The constants,
$a_1$ and $a_2$, measure the effects of a nonuniform density distribution and
clump axial ratio, respectively, on the
gravitational potential and $a_3$ is a statistical correction  to account for 
the projection of an ellipsoidal cloud.  
These constants are evaluated using the functional forms described in 
Bertoldi \& McKee (1992) and with the assumption of 
uniform density clouds ($a_1=1$).  
The value of this combination of constants, ($a_3/a_1a_2$),
is $\sim$1.
The gravitational parameter, 
$$\alpha_G = { {M_{vir}} \over {M_{CO}} }  \eqno(3.2) $$
 provides a measure of the
kinetic to gravitational energy density ratio.  In the outer Galaxy
catalog for which there are only $^{12}$CO observations, 
\mco\ is derived assuming a CO to H$_2$ conversion factor.
Figure~\ref{alphaG} shows the variation of $\alpha_G$ with \lco.
The vertical 
line denotes the
detection limit for L$_{CO}$ at a distance of 10 kpc.
The derived values of the gravitational parameter are anti-correlated 
with the CO luminosity.  A bisector fit to the data yields the 
relationship
$$\alpha_G=53 L_{CO}^{-0.49\pm0.01} \eqno(3.3) $$
with a correlation coefficient of -0.76.
This expression is parameterized in terms of mass assuming 
a constant CO to \htwo\ conversion factor,
$$\alpha_G= \biggl({ {M_\circ}\over{M_{CO}} }\biggr)^{0.49} \eqno(3.4) $$  
where $M_\circ$=1.3$\times$10$^4$ \msun\ and corresponds to the mass
at which $\alpha_G=1$.  
This relationship could be due to the selection 
effect which excludes narrow line regions from the cloud catalog (see $\S2.2$).
The minimum value of the gravitational parameter, $\alpha_G^{min}$, 
is estimated 
by solving for $r_e$ 
in equation 2.7 and inserting the result into equation 2.8 such that,
$$\alpha_G^{min} =  103 \sigma_{vmin}^{3/2} <T>^{-1/2}L_{CO}^{-1/2} \eqno(3.5)$$
where $\sigma_{vmin}$ is the minimum velocity dispersion 
recovered.
For $<T>$=1.4 K and $\sigma_v$=0.43 \kms, 
$$\alpha_G^{min} =  25 L_{CO}^{-1/2} \eqno (3.6) $$
and is shown as the heavy line in Figure~\ref{alphaG}.  The 
functional form of $\alpha_G^{min}$ provides 
a reasonable approximation to the lower envelope of points within the 
$\alpha_G-L_{CO}$ plane.  Therefore, a population 
of clouds with narrow line widths, low luminosities, and smaller
values of $\alpha_G$ could exist within the ISM but is not recovered
in this decomposition.  
Given that there 
are not many points at this limit for a given value of \lco, there may 
not be a significant fraction of clouds with these conditions.
This selection effect is surely present in 
most previous studies of cloud equilibrium.  

The values of \lco\ and $\alpha_G$  are also dependent on the 
assumed kinematic distances to the objects.  The  kinematic 
distances in this sector of the outer Galaxy are often larger
than the spectroscopic distances due to non-circular motions
induced by a large scale potential (Brand \& Blitz 1993).   In these
cases, the derived values of \lco\ and therefore, \mco, are
overestimates to the true values.  Therefore, 
the gravitational parameter is underestimated. 

Finally, the variation of the gravitational parameter with \lco\
can be rectified if the CO to \htwo\ 
conversion factor is not constant but changes systematically
with \lco.  Sodroski (1991) has proposed 
a larger value of 
X$_{CO}$ for the outer Galaxy so that all clouds identified in that 
study are self gravitational.
However, given the strong correlation of $\alpha_G$  over 4
orders of magnitude of \lco, such an ad hoc
modification to X$_{CO}$ implies that
the most luminous objects are collapsing.
Brand \& Wouterloot (1995) examined the variation of the 
conversion factor using \co\ and \coa\ observations for a sample
of far outer Galaxy clouds.  Accounting for a
radial gradient in \coa\ abundance, they 
concluded that the conversion factor is 
similar to that found in the inner Galaxy.  Other \co\ and \coa\
studies have
found no significant variation of $X_{CO}$ for outer Galaxy clouds
(Carpenter, Snell, \& Schloerb 1990).

The conventional assumption of a constant conversion factor is
that clouds are self gravitationally bound with $\alpha_G \approx 1$.
However, as discussed in $\S$2.4, if clouds are 
internally overpressured with respect to 
self gravity ($\alpha_G >>1$), then the appropriate value of $X_{CO}$ 
is smaller than the standard, constant value.  Therefore, by using the 
standard value, the  derived masses are 
upper limits and 
 the derived values
for $\alpha_G$ are lower limits.  
The large values of the gravitational parameter 
reflect the changing dynamical state of molecular 
regions with different mass.  
Only the most luminous objects identified in the Survey 
have sufficient mass to be bound by self gravity ($\alpha_G\sim$1).
Regions with lower CO luminosities and mass are internally
overpressured with respect to self gravity.  This state is 
independent of whether the object is an isolated cloud or part of
a larger cloud complex. 
A similar conclusion has been obtained 
for a sample of high latitude clouds and 
for several clouds in the solar neighborhood
(Magnani, Blitz, \& Mundy 1985;
Keto \& Myers 1986;
Bertoldi \& McKee 1992; Falgarone, Puget, \& Perault 1992; 
Dobashi \etal 1996; Yonekura \etal 1997; Kawamura 
\etal 1998).  The results presented here provide 
statistical confirmation of these earlier studies over 
a larger range of cloud and clump masses.

\subsubsection{Targeted Regions with $^{13}$CO Observations}

In order to gauge the results of the previous section with 
a more reliable tracer of
molecular hydrogen mass, we have analyzed \co\ and 
\coa\ data of targeted molecular cloud regions which lie 
within the Survey field
(Heyer, Carpenter, \& Ladd 1996; Deane 2000).   The targeted fields
include the giant molecular clouds Cep OB3, S140, NGC 7538, and W3.
The \co\ data were decomposed into discrete objects with the same
algorithm as the Survey cube.
The \coa\ integrated 
intensity is summed within the boundaries identified  from the \co\ data 
and a mass, $M_{LTE}$, is derived assuming local
thermodynamic equilibrium, a kinetic temperature of 10 K, 
a \coa\ to \htwo\ abundance of 10$^{-6}$
and a 1.36 correction for the abundance of Helium (Dickman 1978). 
 The distances to each 
cloud are: 730 pc (Cep OB3), 910 pc (S140), 2.35 kpc (W3), and 3.5 kpc 
(NGC 7538).
A virial mass for each object is derived from the tabulated 
size and \coa\
velocity dispersion.

\begin {figure}
\plotone{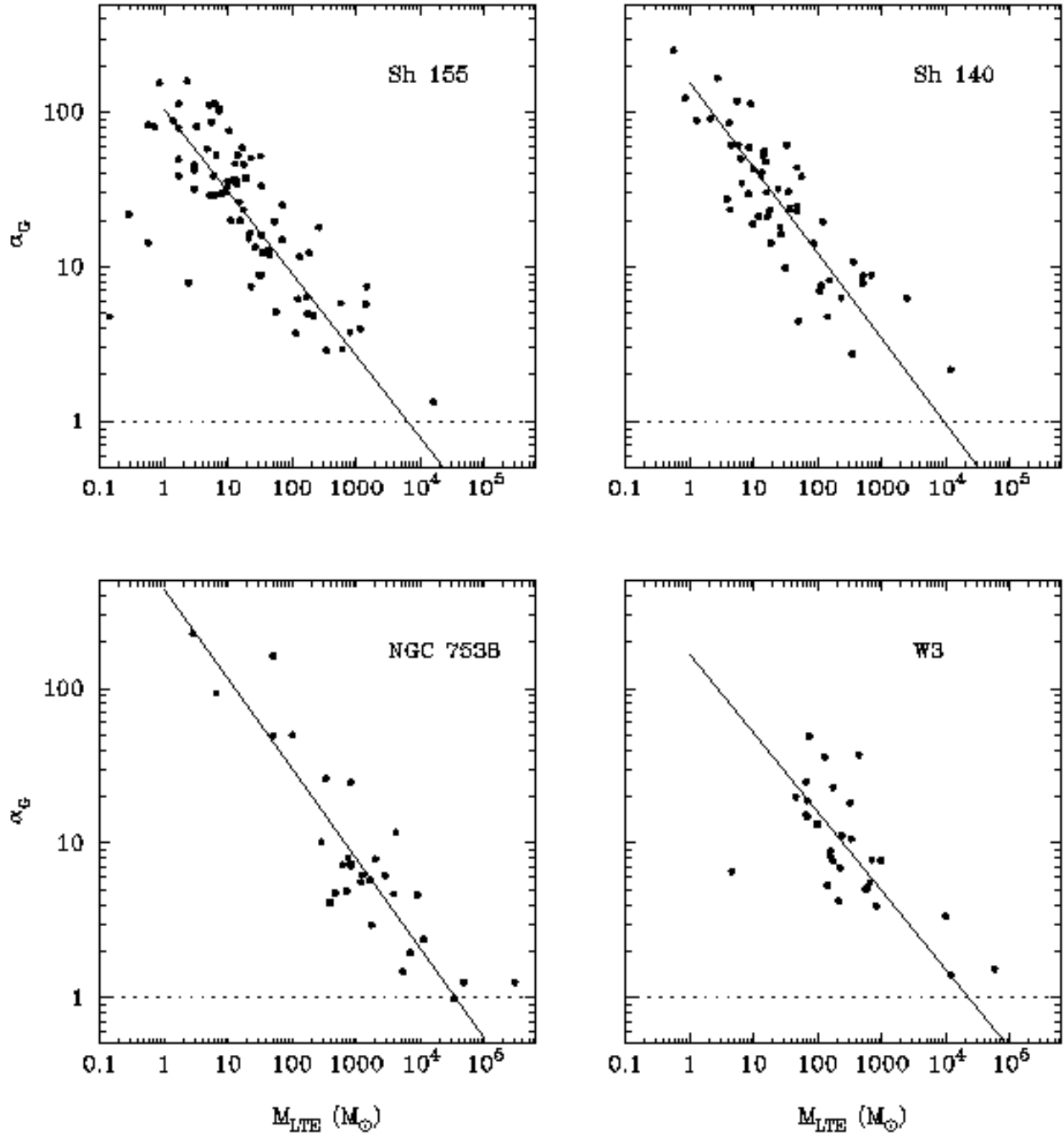}
\caption{The variation of the gravitational parameter, $\alpha_G$,
derived from \coa\ observations for clumps within four 
targeted giant molecular cloud
complexes in the outer Galaxy.
}
\label{alphaG13co}
\end {figure}

The variation of the gravitational parameter, now derived with \coa\
measurements of molecular column density and $M_{LTE}$, is shown in 
Figure~\ref{alphaG13co} for the four targeted giant molecular clouds.  
The evaluation of $\alpha_G$ is subject to the same selection effects
as the Survey clouds such that there is the same functional 
dependence of $\alpha_G^{min}$ with $M_{LTE}$.
$\alpha_G$ decreases with increasing luminosity
and mass.  Bisector fits of the data to the expression
$$\alpha_G = \biggl({ {M_\circ}\over{M_{LTE}} }\biggr)^\epsilon \eqno(3.7) $$
where M$_\circ$ is as defined in equation 3.4, are summarized in Table~2. 
Values of $\epsilon$ range between 0.51 and 0.58.
For objects with masses greater than 10$^4$ \msun, the 
derived values of $\alpha_G$ are reasonably consistent with 
self gravitational equilibrium.  However, the lower mass 
clouds are highly overpressured with respect to self gravity
as found for the Survey clouds.

\begin {figure}
\plotone{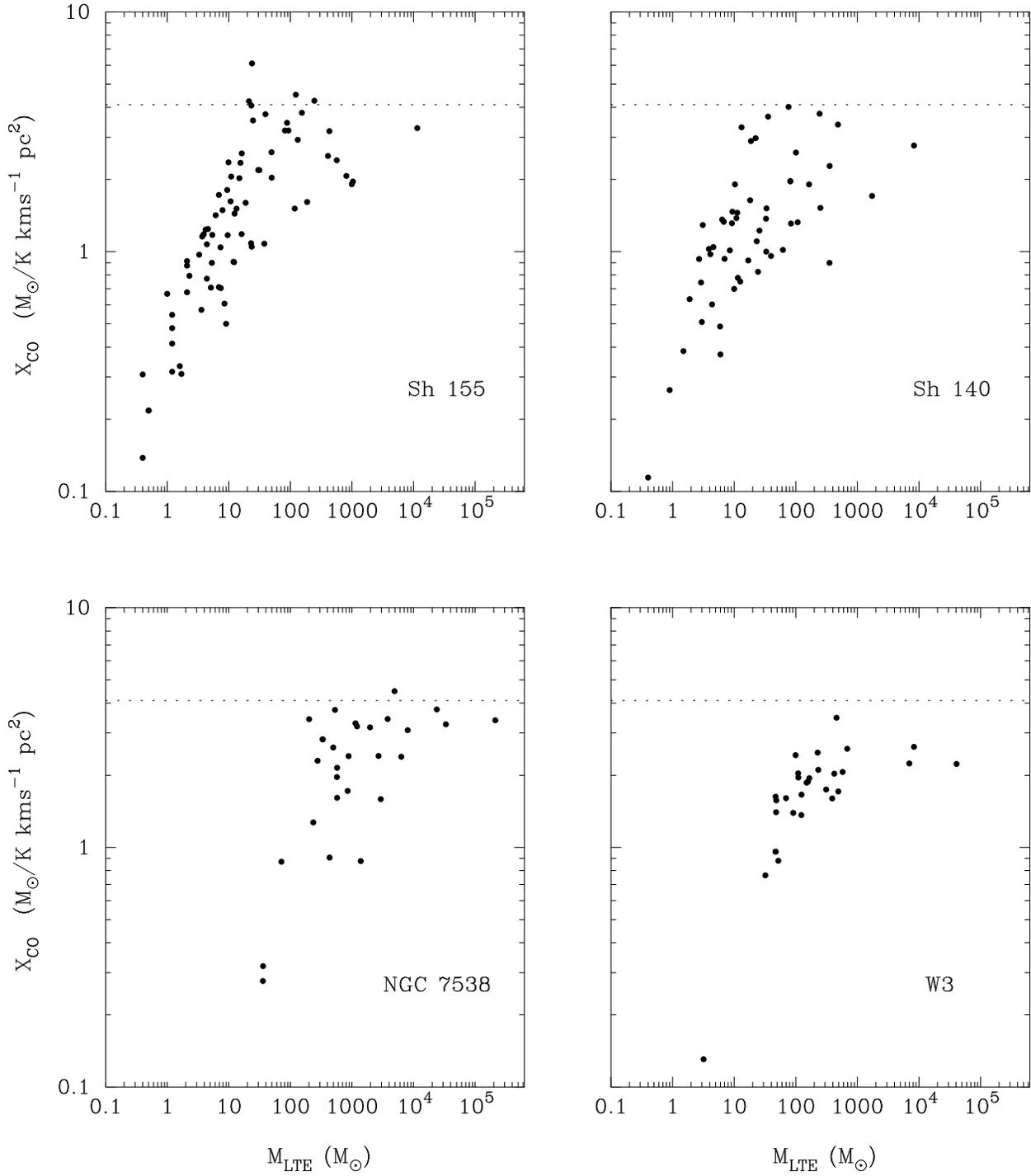}
\caption{The derived CO to \htwo\ conversion factor for clumps
identified within the cloud complexes as a function of mass.
For objects with $\alpha_G{\approx}1$, $X_{CO}$ is 
comparable to the value derived from $\gamma$-ray measurements.
}
\label{XCO}
\end {figure}

Figure~\ref{XCO} shows the 
inferred CO to \htwo\ 
conversion factor,  
$X_{CO}=M_{LTE}/L_{CO}$ as a function of $M_{LTE}$
for the identified clumps within the four fields.
While there are not many  objects with high mass,
the scatter of the ratio for objects with $M_{LTE}>$10$^4$ \msun\
is much smaller than is found for
low mass objects.  The mean value for the high 
mass points is comparable to the value determined from 
$\gamma$-ray measurements (Strong \& Mattox 1996). 
 The inferred values of $X_{CO}$
for the lower mass, non-self-gravitating objects  
are smaller than this standard value as expected from equation 2.7.
These results are direct, empirical evidence 
that 
L$_{CO}$ is a reliable tracer of total mass of
an object for values of 
L$_{CO}$ $>$ 10$^3$ K-km/s pc$^2$ but provides 
only an upper limit to the mass for objects with lower luminosities.

\subsubsection{CO as an Tracer of H$_2$ Mass in Galaxies}
The main isotope of CO provides the primary tracer of H$_2$ mass
in galaxies.
This utility depends upon the 
state of self gravitational equilibrium of the constituent clouds
within the observer's beam (Dickman, Snell, \& Schloerb 1986).  
The non-self gravitating state of 
low luminosity objects does not render the use of CO as an
extragalactic tracer  of \htwo\ mass inapplicable.  Figure~\ref{lco-function}
demonstrates that much of the measured CO luminosity arises from 
the most massive objects within the field for which $\alpha_G \approx 1$.  
Given that \lco\ provides an upper limit to the mass for non-self gravitating
clouds, the mass function 
may be even flatter than the luminosity function.
For most available resolutions of a galaxy, the large, most luminous  
objects would contribute most of the detected flux.  Therefore, if the 
distribution of clouds in other galaxies is similar to the outer Galaxy, then 
$^{12}$CO remains a reliable tracer of \htwo\ mass.  The fractional 
contribution to the measured CO luminosity from the small, non self 
gravitating population of clouds would simply add to the scatter of
inferred molecular hydrogen masses as this contribution could vary 
with position in a galaxy or from one galaxy to another.

\subsection{Surface Densities}
\begin {figure}
\epsscale{0.75}
\plotone{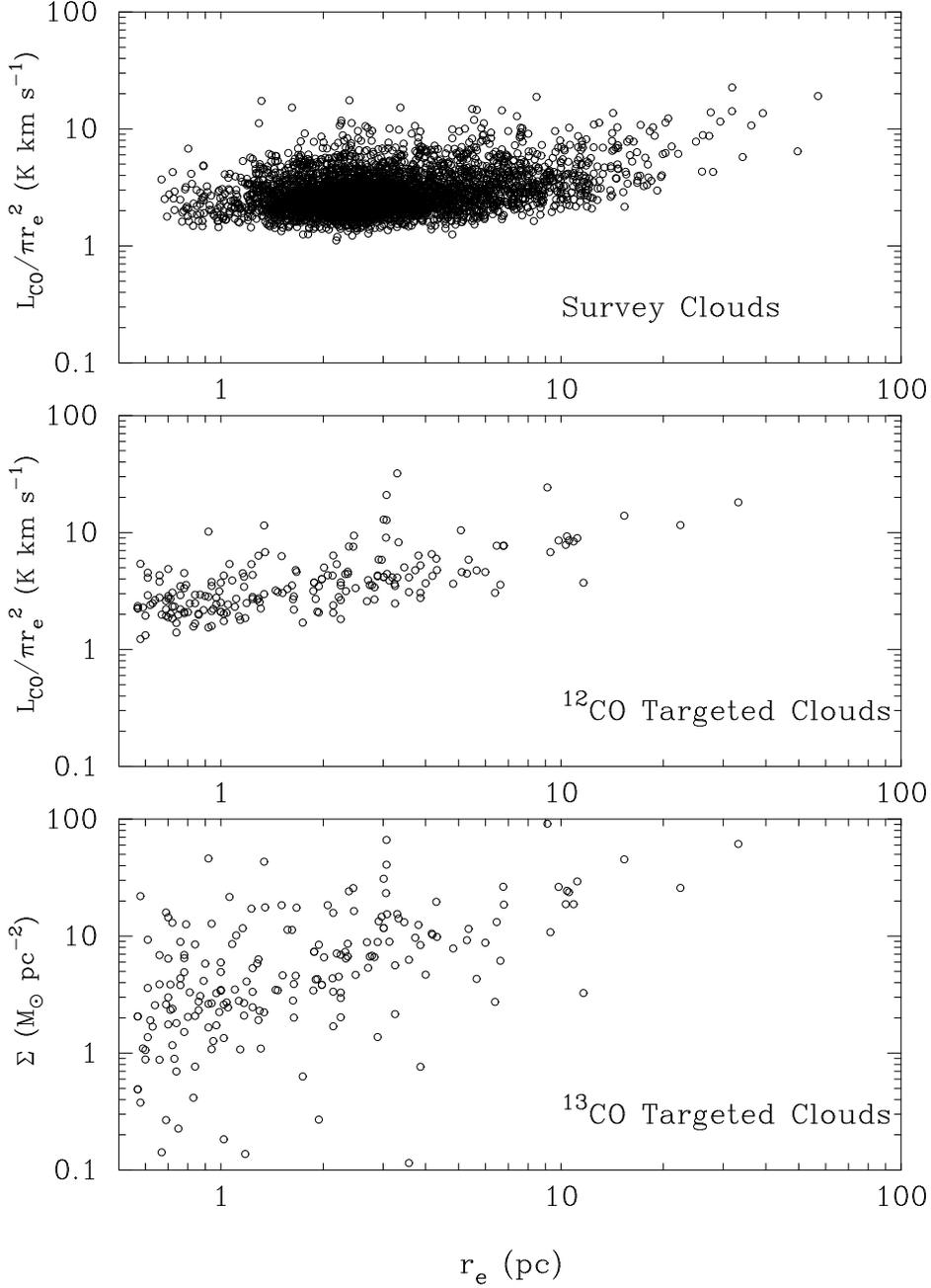}
\caption{(top) 
The variation of mean CO integrated intensity averaged 
over the area of the cloud with the effective cloud radius for the 
sample of Survey clouds.
(middle)
The variation of mean CO integrated intensity averaged 
over the area of the cloud with the effective cloud radius for the 
sample of targeted clouds.
(bottom)
The mass surface density of objects within the targeted clouds
derived from $^{13}$CO observations.
}
\label{sigma}
\end {figure}

The third scaling relationship states that the 
mass surface density of molecular clouds is constant. To evaluate 
this relationship with the Survey clouds, 
the 
variation of CO luminosity normalized to the cloud area (or equivalently,
the mean value of integrated CO intensity) with effective cloud
radius is shown in 
Figure~\ref{sigma}.   A similar relationship is shown from the sample 
of objects identified in the four targeted clouds.
Given the results of $\S$3.2.1, the corresponding
mass surface densities are only valid for the large ($>$10 pc) objects.
There is little variation of the mean integrated intensity for small clouds 
($r_e <$10 pc) 
which also corresponds to the population of non self gravitating clouds.
This is in part due to the method to identify clouds at a given 
threshold of antenna temperature and that most of the luminosity 
arises from the extended, lines of sight with small antenna temperatures.
For these small clouds, \lco\ effectively measures the projected area.
For clouds with effective radii greater than 10 pc, 
there is weak trend for larger mean intensities 
with increasing size.  
The mass surface density of objects is determined from the 
\coa\ observations of the four targeted clouds as shown in Figure~\ref{sigma}.
The mean surface density is 9 M$_\odot$ pc$^{-2}$ although the 
scatter is large for a given size.  It is interesting to note 
that despite the uniformity of mean \co\ intensity between clouds or clumps, 
the mean column density of clumps within a cloud can vary widely.
Thus, the mass surface density is not a constant of molecular clouds.

\section{Discussion}
The preceding sections demonstrate the 
non self gravitating state of molecular regions as defined by \co\
emission with masses
less than 10$^3$ \msun\ while clouds with masses greater than 10$^4$ \msun\ 
are self gravitating.  The limited accuracy
of $\alpha_G$ precludes a definitive evaluation of the equilibrium state 
for regions with masses between 10$^3$ and 10$^4$ \msun\ 
for which $1 <\alpha_G <4$. 
Due to the high opacity of the CO J=1-0 transition, CO observations are
not sensitive to 
the full range of molecular gas column densities known to be present in
clouds.
Therefore, this result does not 
preclude the presence of small, self gravitating regions with 
densities greater than 10$^4$ \cc\ in which star formation may 
occur.  Indeed, star formation is present within many of the small
clouds of 
the W3/4/5 cloud complex 
(Carpenter, Heyer, \& Snell 2000).
Given the luminosity function in Figure~\ref{lco-function}, the small 
cloud and clump populations do not account for a significant fraction of the 
molecular mass in the Galaxy.
Nevertheless, these molecular regions provide insight to the 
dynamical state of the interstellar medium.
The identified 
objects are strictly regions where CO is detected but the 
dominant molecular constituent, \htwo, could extend beyond these
boundaries.  Moreover, these regions are embedded within a 
larger, atomic medium such that the boundaries represent 
a change in gas phase 
rather than sharp volume or column density variations.  

Regions with large values of $\alpha_G$ are either short lived or
are bound by external pressure and long lived with respect to the dynamical 
time scales.
These observations can not distinguish whether a given region
is bound by external pressure as this requires an examination
of the thermal and dynamical state of the surrounding medium.  
In the absence of sufficient external pressure, these regions
expand until the internal pressure is balanced by that of the 
external medium.  In this larger configuration, the molecular gas 
may dissociate due to less effective self-shielding.  Effectively,
the CO emitting regions would be rapidly dispersed over a dynamical 
time along the minimum cloud dimension ($l_{min}/2\sigma_v$), or 
 5-50$\times$10$^4$ years.  
Numerical simulations of magnetohydrodynamic turbulence in the 
dense interstellar medium show localized, density enhancements that would
rapidly 
lose identity due to shear, merging with nearby clouds, or expansion 
into the larger medium  (Ballesteros-Paredes, Vazquez, \& Scalo 1999).  These 
numerical studies
would suggest that the overpressured objects identified in our survey 
are transient features.

\begin {figure}
\plotone{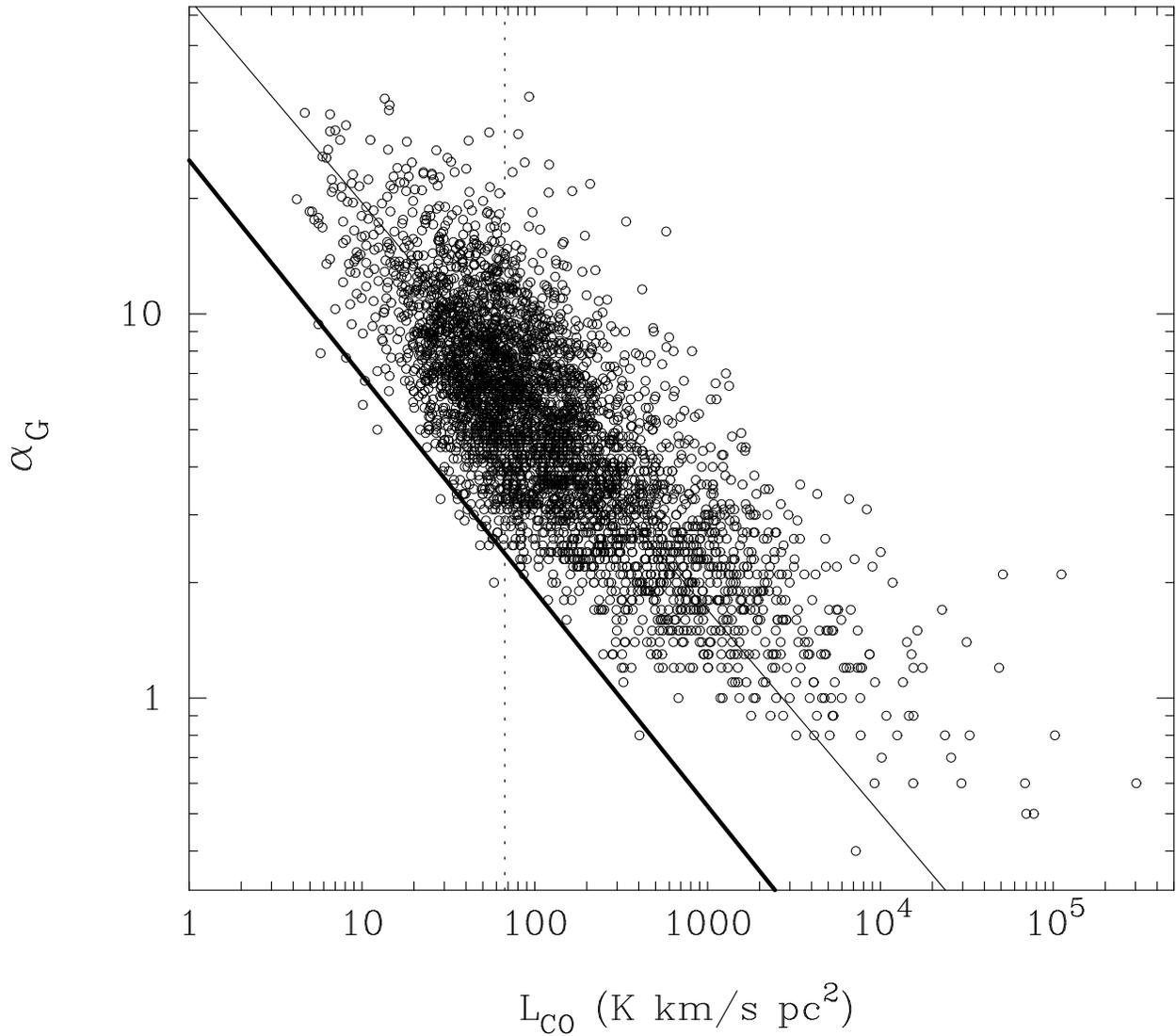}
\caption{The variation of the gravitational parameter, $\alpha_G$,
with  CO luminosity. The dotted vertical line denotes the 
detection limit of \lco\ at a distance of 10 kpc. The heavy line 
shows the minimum value of $\alpha_G$ to which the decomposition is 
sensitive due to the observational selection effect which excludes
narrow line clouds.}
\label{alphaG}
\end {figure}

However, there is indirect evidence to suggest that the 
internal motions of these
 regions are bound by some confining agent.  
The
age of any molecular object is constrained by the time required to
chemically evolve diffuse, atomic material to nominal abundances to
enable a CO observation.  For densities of 10$^2$ \cc, this time scale is 
greater than 10$^6$ years (Jura 1975) which is 
longer than the dynamical time for these objects.  
Unless these overpressured molecular clouds are formed within high density 
regions, there is simply insufficient time to chemically evolve material
to reasonable abundance values.
Secondly, while the lower envelope of points in Figure~\ref{alphaG}
may be defined by selection effects, the upper envelope of values 
of $\alpha_G$ 
systematically varies
through 3 decades of cloud mass (see Figure~\ref{alphaG}).
For transient, unbound objects,
one would expect 
$\alpha_G >> 1$ and independent of mass
such that the observed correlation is unlikely.   
Bertoldi \& McKee (1992) evaluate $\alpha_G$ in terms of the 
Bonner-Ebert mass,
$$ \alpha_G \propto \biggl({ {M_{BE}}\over{M}}\biggr)^{2/3} \eqno(4.1)$$
where
$$ M_{BE} = 1.182 { {\sigma_v^4} \over {(G^3 P_\circ)^{1/2}} } $$
$$ M_{BE} = 
2900 \biggl({{\sigma_v}\over{1 km/s}}\biggr)^4 \biggl({{P_\circ/k}\over{10^4 cm^{-3}K}}\biggr)^{-1/2} 
\eqno(4.2) $$
where $P_\circ$ is the external pressure.
The relationship between $\alpha_G$ and 
{\mco=$X_{CO}$\lco} identified 
in Figure~\ref{alphaG}  is shallower than that predicted assuming 
a constant Bonner-Ebert mass within a given complex as considered by Bertoldi
\& McKee (1992)  
 although this is 
in part, due to a selection effect which excludes regions with small velocity 
dispersions.  
This may also be due to the spatial variation of pressure 
throughout the outer Galaxy 
and the underestimate of the gravitational parameter
for lower mass objects
(see $\S$2.4 and $\S$3.3).
The index is similar to
values derived from \coa\ observations of targeted cloud complexes.  
The results of Dobashi \etal (1996),
Yonekura \etal (1997), and Kawamura \etal (1998) show even 
shallower slopes (0.2-0.3).

\subsection{Required External Pressures}

\begin {figure}
\plotone{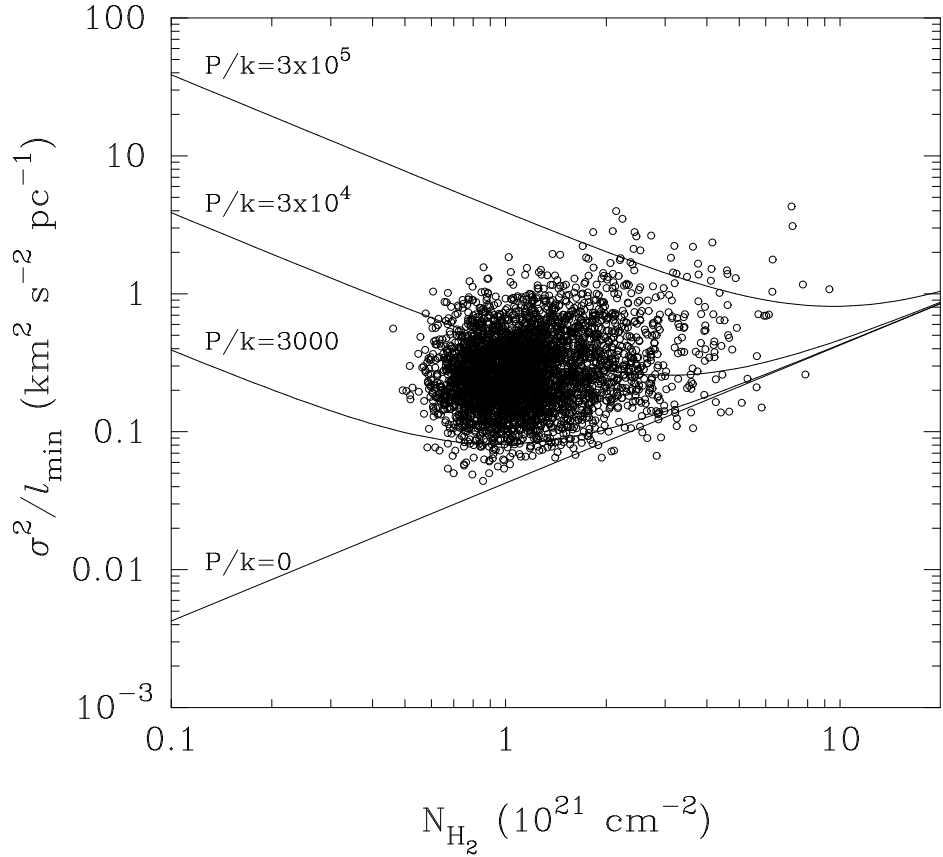}
\caption{ Values of $\sigma_v^2/l_{min}$ vs the mean column density for the 
identified objects.  The solid lines
show the variation of this value for varying 
external pressures for bound objects. }
\label{pext}
\end {figure}

\begin {figure}
\plotone{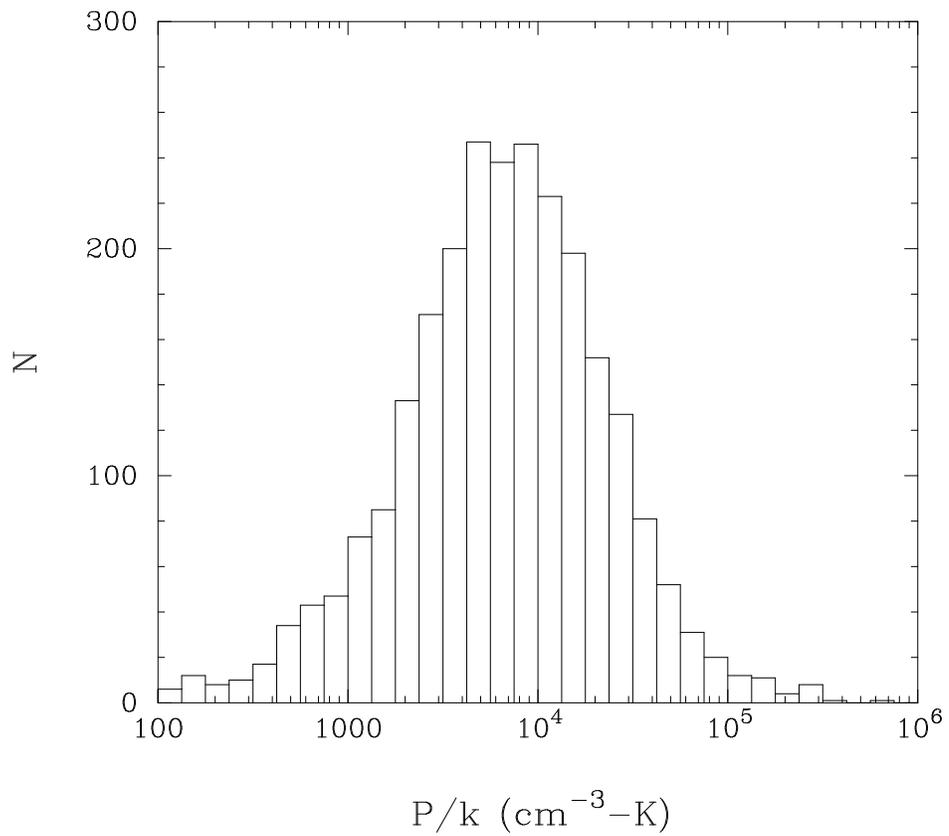}
\caption{ The distribution of required external pressures
to bind the internal motions of identified molecular regions.}
\label{pextdistr}
\end {figure}

The preceding sections demonstrate that there are a large number of
molecular regions whose internal motions are not bound by self 
gravity.   To remain bound in the observed configuration, these 
motions must be confined by the pressure of the external medium.
To gauge the magnitude of the required external pressures, the full
virial theorem
is rewritten, 
$$ {{\sigma_v^2}\over{l_{min}}} = {{P_\circ}\over{k}} 
                {{k}\over{m_{H_2}}} {{1}\over{N_{H_2}}} +
                {{a_1 a_2}\over{a_3}} {{\pi G m_{H_2}}\over{5}} N_{H_2} \eqno(4.3) $$
where \lmin\ is the measured 
minor axis length, $P_\circ$ is the external pressure, 
and $N_{H_2}$ is the mean molecular column density over the surface of the 
object.  This expression assumes that the clouds are prolate. 
Figure~\ref{pext} shows the quantity, $\sigma_v^2/l_{min}$ plotted as a function
of $N_{H_2}$ for each identified object.  Also
shown are the variations of $\sigma_v^2/l_{min}$ with column 
density for different values of the external pressure for 
bound objects.  Self gravitating objects lie along the curve 
P/k=0.  The primary cluster of points lie well off this line.  The 
distribution of required pressure is shown in Figure~\ref{pextdistr}.
The mean and median of the distribution are 1.4$\times$10$^4$
and 6700 cm$^{-3}\;K$ respectively.   No significant variation 
of the required pressure with galactocentric radius can be 
determined given the limited 
dynamic range of $R_{gal}$ and the small number of identified objects
at $R_{gal} >$ 14 kpc.

\subsection{Possible Sources of External Pressures}

Given the magnitude of measured line widths, the internal 
pressure arises from the non thermal, turbulent motions of the gas. 
The required pressures to bind these motions are larger than 
the measured thermal pressures of the interstellar medium
although thermal pressure fluctuations of the required magnitude are observed
within a small fraction of the atomic gas volume 
(Jenkins, Jura, \& Lowenstein 1983; Wannier \etal 1999). 
However, even in the case of comparable external thermal pressure,
the cloud boundary can not be maintained due the anisotropy of 
internal, turbulent gas flow.   An initial perturbation of 
the cloud boundary by a turbulent fluctuation  generates an 
imbalance of the pressure force
perpendicular to the surface which in turn, causes a larger 
distortion of the boundary (Vishniac 1983). 
If the molecular clouds are simply high density
regions resulting from converging gas streams within a larger
turbulent flow, as suggested 
by numerical simulations, there is an effective external ram pressure
component.  However, this component is similarly anisotropic and therefore,
can
not provide the necessary pressure to confine the entire 
boundary of the cloud (Ballesteros-Paredes \etal 1999).  The static 
magnetic field applies an effective pressure, $B^2/8{\pi}k$, to 
the molecular gas and may contribute to the pressure support of the 
cloud.  For a 5$\mu$gauss field, this pressure is 7200
$cm^{-3}\;K$ and comparable to the values required to bind the 
non-self-gravitating clouds observed in this study.

Bertoldi \& McKee (1992) propose
that the weight of the self-gravitating cloud complex squeezes the 
interclump medium to 
provide an effective mean pressure upon a constituent clump.  
The magnitude of this 
pressure, $<P_G>$, is equivalent to the gravitational energy density of the 
cloud complex,  $G M_c^2/R_c^4$
where $M_c$ and $R_c$ are the mass and radius of the cloud complex 
respectively.
They demonstrate that the magnitude of this pressure is 
similar to the required pressure to bind the clumps within the 
four targeted cloud complexes which they analyzed.  
If the weight of the molecular complex is a 
significant component to the 
equilibrium of clumps, then the required pressure
should vary with location of the clump within the 
gravitational potential.  
In this study, the identified objects are not grouped into cloud complexes 
and therefore,
the self gravity of the larger complex is not evaluated.

In the outer Galaxy, the surface density of atomic gas is much larger 
than that of the molecular material and therefore, 
provides an additional component to 
the weight upon a given clump or isolated cloud. 
The mean effective external 
pressure at the molecular gas boundary,  
due to the 
overlying atomic gas layer is
$$ <P_G/k> = P_e/k + (G/k) ({N_H \mu m_H})^2 \eqno(4.4) $$
where $P_e/k$ is the kinematic pressure at the external atomic gas boundary,
and $N_H \mu m_H$ is the mass surface density of the atomic gas across the 
disk
(Elmegreen 1989).  The kinematic pressure at the atomic gas 
boundary  is estimated from the 
respective surface density and velocity dispersion 
of gas and stars to be 8000 cm$^{-3}$ K (Elmegreen 1989).   
To self shield the molecular gas, the column density of atomic 
gas in the near vicinity 
of molecular material is 2$\times$10$^{21}$ cm$^{-2}$ 
 such that the 
$ <P_G/k> \approx 1.8{\times}10^4$ cm$^{-3}$-K which is comparable to 
the magnitude of pressures shown in Figure~\ref{pextdistr}.  Therefore, it 
is plausible that the weight of the HI layer of gas or magnetic fields 
provide the 
external pressure to maintain equilibrium of the low mass molecular 
regions in the outer Galaxy.

\section{Conclusions}
A decomposition of the FCRAO CO Survey of the outer Galaxy has 
identified 10,156
discrete regions of molecular gas.  A subset of this 
catalog is analyzed with $V_{LSR} < -20$ \kms\ which includes objects within the
Perseus arm and far outer Galaxy. 
\begin{enumerate}
\item Molecular regions with masses less than 10$^3$ \msun\ are not self
gravitational.  For these regions, \htwo\ masses derived from
a CO to \htwo\ conversion factor are upper limits.
\item The pressures required to bind the internal motions of these 
non self gravitating regions
are 1-2$\times$10$^4$ cm$^{-3}$-K. 
The
weight of the atomic gas layer in the disk may provide this
necessary pressure to maintain the equilibrium of these clouds.
\item The \co\ luminosity function, ${\Delta}N/{\Delta}L_{CO}$, varies as a power law,
$$ {\Delta}N/{\Delta}L_{CO} \propto L_{CO}^{-1.80\pm0.04}.$$  However, given the 
non self gravitating state of low luminosity clouds, this relationship
should not be used to infer a mass spectrum of molecular clouds.
\item The \co\ velocity dispersion of a cloud is invariant with the size 
for clouds with radii less than 7 pc.
\end{enumerate}

We acknowledge valuable discussions with Enrique Vazquez-Semadeni,
Javier Ballesteros-Paredes and Jonathon Williams.
This work is supported by
NSF grant AST 97-25951 to the Five College Radio Astronomy
Observatory.

\clearpage
\section{References}
\begin{description}
\item[] Ballesteros-Paredes, J., Vazquez-Semadeni, E., \& Scalo, J. 1999, ApJ, 
515, 286
\item[] Bally, J., Stark, A.A., Wilson, W.D., \& Henkel, C.  1988, ApJ, 324, 223
\item[] Bertoldi, F. \& McKee, C.F. 1992, ApJ, 395, 140
\item[] Blitz, L. \& Spergel, D.N. 1991, ApJ, 370, 205
\item[] Brand, J. \& Blitz, L. 1993, AA, 275, 67
\item[] Brand, J. \& Wouterloot, J.G.A. 1994, AAS, 103, 503
\item[] Brand, J. \& Wouterloot, J.G.A. 1995, AA, 303, 851
ApJ, 324, 248
\item[] Carpenter, J.M., Snell, R.L., \& Schloerb, F.P. 1990, ApJ, 362, 147
\item[] Carpenter, J.M., Snell, R.L., \& Schloerb, F.P. 1995, ApJ, 445, 246
\item[] Carpenter, J.M., Heyer, M.H., \& Snell, R.L. 2000, ApJ, in press
\item[] Clemens, D.P. \& Barvainis, R. 1988, ApJS, 68, 257
\item[] Dame, T. M., Ungerechts, H., Cohen, R. S., de Geus, E. J., Grenier,
I. A., May, J., Murphy, D. C., Nyman, L. A., \& Thaddeus, P. 
1987, ApJ, 322, 706
\item[] Deane, J. 2000, Ph.D dissertation, University of Hawaii
\item[] Dobashi, K. Bernard, J.P. \& Fukui, Y. 1996, ApJ, 466, 282
\item[] Dickman, R.L. 1978, ApJS, 37, 407
\item[] Dickman, R.L., Snell, R.L., \& Schloerb, F.P. 1986, ApJ, 309, 326
\item[] Elmegreen, B.G. \& Lada, C.J. 1977, ApJ, 214, 725
\item[] Elmegreen, B.G. 1989, ApJ, 338, 178
\item[] Elmegreen, B.G. \& Falgarone, E. 1996, ApJ, 471, 816
\item[] Falgarone, E., Puget, J.-L., \& Perault, M. 1992, AA, 275, 715
\item[] Falgarone, E. 1996, in Starbursts, Triggers, Nature, and Evolution
eds. B. Guiderdoni \& A. Kembhavi (Springer:Berlin), p. 41
\item[] Goodman, A.A., Barranco, J.A., Wilner, D.J., \& Heyer, M.H.  1998, 
ApJ, 504, 223
\item[] Heyer, M.H., Carpenter, J., \& Ladd, E.F. 1996, ApJ, 463, 630 
\item[] Heyer, M.H., Brunt, C., Snell, R.L., Howe, J., Schloerb, F.P., \& 
Carpenter, J.M. 1998, ApS, 115, 241
\item[] Hildebrand, R.H., 1983, QJRAS, 24, 267 
\item[] Jenkins, E.B., Jura, M., \& Lowenstein, M. 1983, ApJ, 270, 88
\item[] Jura, M. 1975, ApJ, 197, 575
\item[] Kawamura, A. Onishi, T., Yonekura,Y., Dobashi, K., 
Mizuno, A., Ogawa, H., \& Fukui, Y. 1998, ApJS, 117, 387
\item[] Keto, E. \& Myers, P.C. 1986, ApJ, 304, 466
\item[] Kramer, C., Stutzki, J., Rohrig, R., \& Corneliussen, U.
1998, AA, 329, 249
\item[] Kwan, J. 1979, ApJ, 229, 567
\item[] Larson, R.B. 1981, MNRAS, 194, 809
\item[] Lizst, H.S. \& Burton, W.B.  1981, ApJ, 243, 778
\item[] Magnani, L., Blitz, L., \& Mundy, L. 1985, ApJ, 295, 402
\item[] Myers, P.C. 1983, ApJ, 270, 105
\item[] Sanders, D. B., Scoville, N.Z., \& Solomon, P.M.
1985, ApJ, 289, 373
\item[] Sanders, D. B., Clemens, D. P., Scoville, N. Z., \& Solomon, P. M.
1985, ApJS, 60, 1
\item Scalo, J. 1990, in Physical Processes in Fragmentation and Star Formation,
eds. Capuzzo-Doletta R. \etal Reidel:Dordrecht, p. 151
\item[] Scoville, N.Z., Yun, M.S., Sanders, D. B., Clemens, D. P., 
\& Waller, W.H. 1987, ApJ,S, 63, 821
\item[] Sodroski, T.J. 1991, ApJ, 366, 95
\item Solomon, P.M., Rivolo, A.R., Barrett, J., \& Yahil, A. 1987, ApJ, 
319, 730
\item[] Strong, A.W. \& Mattox, J.R. 1996, AA, 308, L21
\item[] Stutzki, J., Bensch, F., Heithausen, A., Ossenkopf, V., \&
Zielinsky, M. 1998, AA, 336, 697
\item[] van Dishoeck, E. \& Black, J. H., 1988, ApJ, 334, 771
\item[] Vishniac, E.T. 1983, ApJ, 274, 152
\item Wannier, P, Andersson, B-G., Penprase, B.E., \& Federman, S.R.
1999, ApJ, 510, 291
\item[] Yonekura,Y., Dobashi, K.,
Mizuno, A., Ogawa, H., \& Fukui, Y. 1997, ApJS, 110, 21
\item[] Zinnecker, H., McCaughrean, M.J., Wilking, B.A. 1993,
in Protostars \& Planets III, eds.   429
\end{description}

\clearpage 
\appendix
\section {Description of Object Parameters}
Discrete objects are identified within the $T(l,b,v)$ data cube 
as a closed surface such that all values are greater than or equal to 
a singular threshold of antenna temperature.  The three dimensional 
pixels (hereafter, voxels) are contiguous within the volume.  For a
given angular position, ($l_i,b_i$), a minimum of 2 contiguous 
channels are required to exceed the threshold.
In practice, the program finds a seed voxel
above the threshold and then recursively checks neighboring channels and 
positions to build up an object.  Once checked, the voxel is flagged
so it would not be checked again.  A minimum of 5 angular pixels are
required for an object to be included in the final catalog.  

An object is comprised of N angular pixels with each pixel, i,
contributing $P_i$ spectroscopic channels.
A centroid position ($l_c,b_c,v_c$) is calculated from the intensity
weighted mean position within the image,
$$ l_c = { {\sum_{i=1}^N \sum_{j=1}^{P_i} T(l_i,b_i,v_j)l_i } \over 
        { \sum_{i=1}^N \sum_{j=1}^{P_i} T(l_i,b_i,v_j) } }  \eqno(A1.1) $$
$$ b_c = { {\sum_{i=1}^N \sum_{j=1}^{P_i} T(l_i,b_i,v_j)b_i } \over 
        { \sum_{i=1}^N \sum_{j=1}^{P_i} T(l_i,b_i,v_j) } } \eqno(A1.2) $$
$$ v_c = { {\sum_{i=1}^N \sum_{j=1}^{P_i} T(l_i,b_i,v_j)v_j } \over 
        { \sum_{i=1}^N \sum_{j=1}^{P_i} T(l_i,b_i,v_j) } }  \eqno(A1.3) $$
A kinematic distance, D, to the object is obtained by assuming a flat rotation 
curve
$$ D =  R_\circ { {(cos l_c+\sqrt{cos^2 l_c - (x^2-1)})} \over {cos b_c} } \eqno(A1.6) $$ 
where
$$ x = {{1}\over{1+v_c/\Theta_\circ sin l_c cos b_c} } $$
The associated galactocentric radius, R, and  scale height, z, are
$$ R_{gal} = x R_\circ $$
$$ z = D sin(b_c) $$

To parameterize the internal motions within an object, we calculate 
the equivalent width from the composite spectrum of the object.
$$ {\delta}v = 
        \sum_{j=P_{min}}^{P_{max}} \psi(v_j) dv / MAX(\psi(v_j)) \eqno(A1.7) $$
where
$$ \psi(v_j) =  \sum_{i=1}^N T(l_i,b_i,v_j) $$
and $P_{min}$ and $P_{max}$ are the minimum and maximum spectroscopic 
contributing channels over all the pixels. 
${\delta}v$ is an approximation to the full width half maximum line 
width of a centrally peaked spectrum.  It includes motions
along the line of sight as measured by the width of individual line
profiles and more macroscopic motions from the variations of 
the centroid velocity over the projected surface of the object.
This latter component is also tabulated directly,
$$ {\delta}v_c = \sqrt{8ln2}*\sqrt{\sum_{i=1}^N (V_{LSR}^i-v_c)^2/(N-1)} \eqno(A1.8) $$

Given the complex distribution of \co\ emission, we have made 
estimates to object sizes, axial ratios, and orientations from 
simple measures of the associated pixels which comprise an object.  
The 
long dimension, \lmax, 
of an object is determined from the two vertices, $(l_i,b_i), (x_j,y_j)$,
with the 
largest angular separation,
$$ l_{max} = 10^3 \Omega^{1/2} D (MAX(\sqrt{(l_i-x_j)^2+(b_i-y_j)^2}))\;\;\ pc  \eqno(A1.9) $$
A mean, minimum distance, \lmin, is derived such that,
$$ \pi l_{max} l_{min} / 4 = 10^6 N \Omega D^2 \;\;\; pc^2 \eqno(A1.10) $$
The orientation of the cloud within the Galaxy, $\theta$, is the angle of the 
major axis with respect to the positive latitude axis, measured clockwise.  
Obviously, this angle is only meaningful
for those objects with large axial ratios, \lmax/\lmin.

The CO luminosity, \lco, is calculated directly from the associated voxels
and the kinematic distance to the object,
$$ L_{CO} = 10^6 \Omega D^2 {\delta}v \sum_{i=1}^N \sum_{j=1}^{P_i} T(l_i,b_i,v_j) 
\;\;\; K\; km\; s^{-1}\;pc^2 \eqno(A1.11) $$  

Finally, the peak antenna temperature from all of the associated 
voxels is tabulated, 
$$ T_p = MAX [T(l_i,b_i,v_j)] \;\;\; for\;\; i=1,N \;\;and\;\; j=1,P_i 
\eqno(A1.12) $$
\section{Recovery of Line Widths}

To gauge the accuracy and limitations of the derived 
line widths of an object, we have applied the 
cloud decomposition algorithm to a set of model 
clouds with varying velocity dispersion and signal to noise.
Each model cloud is described by a gaussian distribution
along the angular and spectroscopic coordinates an amplitude, T$_\circ$
such that 
$$ T(x,y,v)=T_\circ exp(-2.77(x/R_x)^2)exp(-2.77(y/R_y)^2)
exp(-2.77(v/{\delta}v_t)^2) + T_N  $$
where $R_x$ and $R_y$ are the full width half maximum sizes of the 
cloud, ${\delta}v_t$ is the full width half maximum line width, 
and $T_N$ is extracted from a distribution of gaussian noise 
with variance, $\sigma^2$. 
The size of the model cloud is fixed with $R_x=R_y=5$ pixels.
The centroid velocity is held constant with respect to 
the angular coordinates so all of the  velocity dispersion is 
due to the intrinisic line width of the line profiles.
Models are generated with ${\delta}v_t$
ranging from 1.0 to 3.5 \kms\ and varying signal to 
noise.  
Ten realizations for each cloud are constructed 
with a different noise field generated from a new random number seed.

\begin {figure}
\plotone{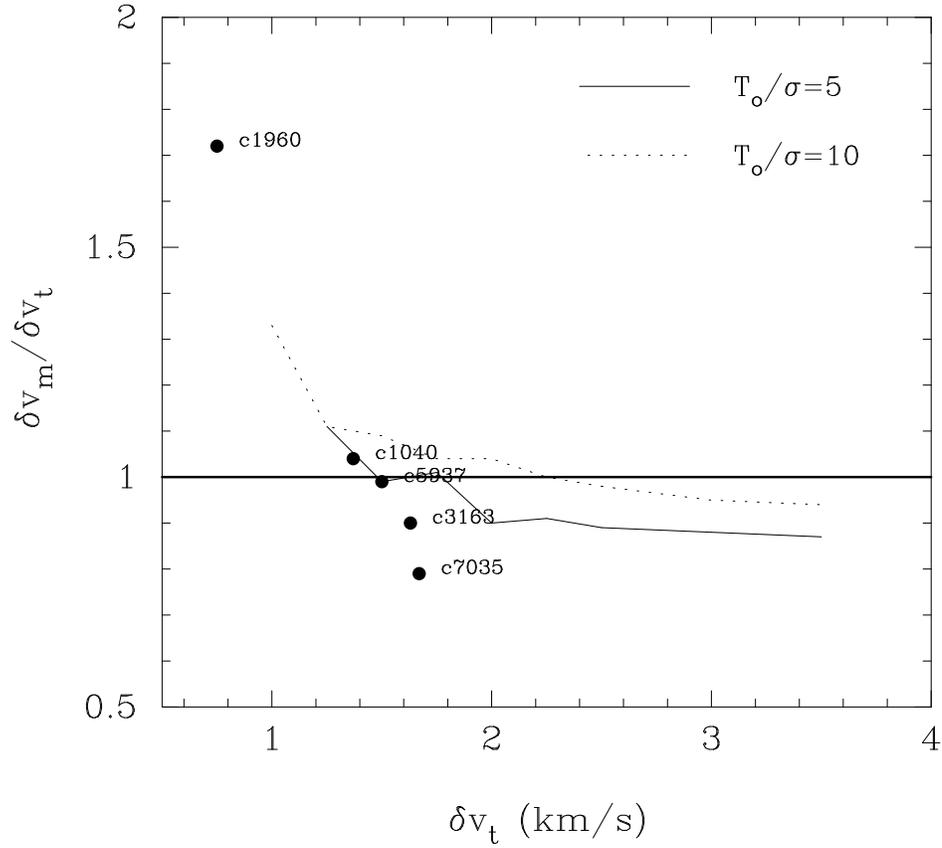}
\caption{ 
The variation of the ratio of measured to intrinsic line 
widths as a function of the intrinsic line width derived from a
sample of model clouds with signal to noise ratios of 5 (solid line)
and 10 (dashed line).  The filled circles show this ratio derived 
from high resolution, high sampling observations of four Survey 
clouds.
}
\label{a1}
\end {figure}

The decomposition program is applied to this model data cube with 
a threshold of 1.5$\sigma$ as in the analysis of the Survey.  
Figure~\ref{a1} shows the variation of the ratio of measured 
full width half maxium line to the intrinsic line width
as a function of the intrinsic line width for models with signal to noise
ratios of 5 and 10.
For intrinsic line widths greater than 1.5 \kms, this ratio 
approaches unity with increasing signal to noise.   The measured 
values are underestimated by $\sim$15\%.  For intrinsic line widths 
of 1.5 \kms, the measured values are overestimated by 10\% and do not 
appear to asymtotically approach the intrinsic widths for increasing 
signal to noise.  Finally, for intrinsic widths of 1.0 \kms,
the spectra are not resolved by the spectrometer.  Objects with 
such narrow line widths can only 
be identified for signal to noise ratios greater than 6.
The velocity dispersions are well determined for 
regions with peak antenna temperatures greater than 4$\sigma$
and intrinsic FWHM line widths greater than 1.5 \kms.

As a secondary measure to the accuracy to which 
the velocity dispersion is recovered, 
we have reobserved 
several small clouds identified in the cloud catalog with 
much higher spectral resolution (0.05 \kms) and angular sampling
($\sim$22\arcsec).  The clouds were selected to be small and to 
cover a limited range of line widths, ${\delta}v_m$. 
For each new map, a line
width is calculated from the average spectrum and is presumed to 
provide an accurate measure of the true line width, ${\delta}v_t$
for each cloud.  
The results are shown as the solid circles in 
Figure~\ref{a1}.  While limited in number, these  measurements 
confirm our ability to recover the line widths for 
${\delta}v >$ 1.4 \kms.


\clearpage
\begin{table}[htb]
\begin{center}
\caption{Derived Properties of Identified Objects}
\vspace{7mm}
 \begin{tabular}
{l}
\hline
This table is available only on-line as a machine-readable table\\
\hline
 \end{tabular}
 \end{center}
\end{table}

\begin{table}[htb]
\begin{center}
\caption{Fitted Parameters to $\alpha_G=(M_\circ/M)^\epsilon$}
\vspace{7mm}
 \begin{tabular}
{lcc}
\hline
Complex & $\epsilon$ & $M_\circ$ \\
        &      & (M$_\odot$) \\
\hline
Sh 140 & 0.55 $\pm 0.03$ & 9500 \\
Cep OB3 & 0.53 $\pm 0.01$ & 6500 \\
W3     & 0.51 $\pm 0.01$ & 2.3$\times$10$^4$ \\
NGC 7538 & 0.58 $\pm 0.04$ & 3.6$\times$10$^4$ \\
Survey Clouds & 0.49 $\pm 0.03$ & 1.1$\times$10$^4$ \\
\hline
 \end{tabular}
 \end{center}
\end{table}
\end{document}